\documentclass{article}
\usepackage{amssymb,amsmath}
\newcounter{mnote}
\newtheorem{lemma}{Lemma}
\newtheorem{proposition}{Proposition}
\newtheorem{theorem}{Theorem}

\newcommand{\proof}{\noindent {\bf Proof. }}
\newcommand{\proofof}[1]{\noindent {\bf Proof of #1. }}
\newcommand{\qed}{\hfill $\fbox{\hspace{0.3mm}}$ \vspace{.3cm}} 

\newcommand{\Natural}{\mathbb N}
\newcommand{\Real}{\mathbb R}
\newcommand{\Complex}{\mathbb C}
\newcommand{\id}{\mbox{id}}
\newcommand{\ran}{\mbox{Ran}}

\newcommand{\hateq}{\; \hat{=}\; }

\newcommand{\grad}{{\bf grad}\,}    
\newcommand{\curl}{{\bf curl}\,}    
\newcommand{\dvrg}{{\bf div}\,}     
\newcommand{\M}{{\cal M}}           
\newcommand{\Ham}{{\cal H}\,}       
\newcommand{\Ric}{Ric\,}            
\newcommand{\lie}{{\cal L}}         

\title{A minimization problem for the lapse and the initial-boundary
value problem for Einstein's field equations}

\date{\today}
\author{Gabriel Nagy$^1$ and Olivier Sarbach$^{1,2}$\\[0.2cm]
$^1$Department of Mathematics,\\ 
University of California at San Diego,
9500 Gilman Drive,\\
La Jolla, California 92093--0112, USA\\[0.2cm]
$^2$Theoretical Astrophysics 130-33,\\
California Institute of Technology,\\
Pasadena, California 91125, USA}

\begin{document}
\maketitle

\begin{abstract}
We discuss the initial-boundary value problem of General Relativity.
Previous considerations for a toy model problem in electrodynamics
motivate the introduction of a variational principle for the lapse
with several attractive properties. In particular, it is argued that
the resulting elliptic gauge condition for the lapse together with a
suitable condition for the shift and constraint-preserving boundary
conditions controlling the Weyl scalar $\Psi_0$ are expected to yield
a well posed initial-boundary value problem for metric formulations of
Einstein's field equations which are commonly used in numerical
relativity.

To present a simple and explicit example we consider the $3+1$
decomposition introduced by York of the field equations on a cubic
domain with two periodic directions and prove in the weak field limit
that our gauge condition for the lapse and our boundary conditions
lead to a well posed problem. The method discussed here is quite
general and should also yield well posed problems for different ways
of writing the evolution equations, including first order symmetric
hyperbolic or mixed first-order second-order formulations. Well posed
initial-boundary value formulations for the linearization about
arbitrary stationary configurations will be presented elsewhere.
\end{abstract}

\section{Introduction}

A common approach to numerical relativity is to find solutions of
Einstein's field equations on a cylinder of the form $I \times
\Sigma$, where $I = [0,T]$ is a time interval and $\Sigma$ is a
three-dimensional compact manifold with smooth boundary
$\partial\Sigma$. The initial-boundary value problem (IBVP) consists
in finding solutions with specified (initial) data on the ``bottom''
of this cylinder (i.e. on $\{ 0 \} \times \Sigma$) and boundary data
on the set ${\cal T} \equiv I \times \partial\Sigma$. The exact form
of the boundary data depends on the nature of ${\cal T}$; in this
article we consider the case where ${\cal T}$ is timelike. This
excludes the case of inner black hole excision boundary conditions
where ${\cal T}$ is spacelike or null, and the case where spacetime is
compactified in such a way that ${\cal T}$ is a null surface.

For timelike ${\cal T}$, boundary conditions must be specified with
care. First, they should be stable in the mathematical sense of well
posedness of the resulting IBVP which means that one can guarantee
local in time existence of solutions which lie in an appropriate
normed function space and which depend uniquely and continuously on
the initial and boundary data. Second, the boundary conditions should
be constraint-preserving which means that they must guarantee that a
solution to the IBVP with constraint-satisfying initial data
automatically satisfies the constraints everywhere on $I \times
\Sigma$. Third, the boundary data should enable control over the
gravitational flux through the boundary, at least in some approximate
sense. This is a difficult question since in General Relativity there
are no meaningful local expressions for quantities representing the
energy density of the gravitational field. This is related to the fact
that locally, the gravitational field can always be transformed
away. Finally, it would be desirable that boundary conditions could be
specified in such a way to control part of the geometry of ${\cal T}$.

A well posed IBVP of Einstein's vacuum field equations addressing all
the above points has been presented in \cite{hFgN99}. This work is
based on the use of tetrad fields in a gauge adapted to the boundary
${\cal T}$ and casting the evolution equations into a first order
symmetric hyperbolic system for the tetrad fields, the connection
coefficients and the components of the Weyl tensor. Furthermore, the
evolution equations in \cite{hFgN99} imply that the constraint
propagation system, describing the evolution of the constraint fields,
has the form of a hyperbolic system which on ${\cal T}$ contains only
derivatives tangential to ${\cal T}$. As a consequence, the
satisfaction of the constraint-preserving property is automatic. The
only fields with non-trivial characteristic speeds at the boundary are
the Weyl scalars $\Psi_0$ and $\Psi_4$ that can be interpreted as
describing in- and outgoing gravitational radiation, at least in some
approximate sense. (Notice that our definition of $\Psi_0$ and
$\Psi_4$ differs from the one in \cite{hFgN99}.) By specifying data to
$\Psi_0$ or a suitable combination of $\Psi_0$ and $\Psi_4$, the work
in Ref. \cite{hFgN99} obtains a well posed IBVP for Einstein's vacuum
field equations by applying standard theorems on symmetric hyperbolic
equations with maximally dissipative boundary conditions
\cite{jR85,pS96a,pS96b}.

On the other hand, most numerical codes are based on formulations of
Einstein's equations which use metric variables and not tetrad fields.
For such formulations, the problem of obtaining a well posed IBVP is
open. One of the major obstacles stems from the fact that
compatibility with the constraints yields boundary conditions which
have the form of differential equations on the boundary for which no
standard mathematical theorems are known. Despite several partial
results
\cite{bSbSjW02,bSjW03,mBbSjW06,gCjPoRoSmT03,cGjM04b,nT-PhD-05}, the
derivation of necessary conditions for well posedness
\cite{jS98,gCoS03,oSmT05,cGjM04b,oR-PhD-05} and numerical experiments
\cite{gClLmT02,jBlB02,bSbSjW02,bSjW03,oSmT05,lKlLmSlBhP05,gC-PhD-03,cBtLcPmZ05,lLmSlKRrOoR05,mBbSjW06},
a full understanding of the IBVP for metric formulations has not yet
been achieved.

In this article, we reconsider the IBVP for Einstein's field equations
by combining the ideas of Ref. \cite{hFgN99} with new ideas developed
in \cite{oRoS05} for a model problem in electrodynamics. These new
ideas, which are reviewed in Sect. \ref{Sect:MainIdeas}, are based on
separate estimates for the constraint fields, the curvature fields and
the imposition of an appropriate gauge condition which allows an
estimate for the remaining, gauge-dependent fields. Boundary
conditions on the constraint and curvature fields are an essential
part in obtaining these estimates. In the electromagnetic case, the
curvature fields are the electric and magnetic fields which are
gauge-invariant. In General Relativity, the curvature field is
described by the Riemann tensor which can be decomposed into the Ricci
tensor and the Weyl tensor. Motivated by the model problem in
\cite{oRoS05} we consider a gauge condition for the lapse that is
obtained by minimizing a simple functional representing the norm
squared of the time derivative of the extrinsic curvature. The
minimization principle, which yields a fourth order elliptic boundary
value problem for the lapse, possesses several useful properties. One
of them is that the functional is positive semi-definite and zero if
and only if the components of the extrinsic curvature are
time-independent. Therefore, the gauge condition should be well
adapted to stationary configurations or small deviations thereof. In
Sect. \ref{SubSect:IBVP} we summarize the boundary conditions, the
gauge conditions and the evolution system considered in this article
and write down the resulting IBVP. In Sect. \ref{Sect:LapseGauge} we
examine the variational principle for the lapse in the weak field
limit, establish its well posedness in Theorem \ref{Thm:Min}, and give
a physical interpretation for it as a gauge-fixing procedure for the
extrinsic curvature. In Lemma \ref{Lem:KeyEstimate} we also derive
some key estimates for the extrinsic curvature which are valid in this
gauge.

The main result of this article is derived in Sect. \ref{Sect:WP}.
There, we consider the linearized Einstein vacuum field equations in
the $3+1$ decomposition as described in Ref. \cite{jY79} and couple
them to the fourth order elliptic gauge condition for the lapse. We
prove in Theorem \ref{Thm:Main} that this elliptic-hyperbolic system,
together with constraint-preserving boundary conditions controlling a
suitable combination of the Weyl scalars $\Psi_0$ and $\Psi_4$ leads
to a well posed IBVP. The proof is based on casting the evolution
system into an abstract Cauchy problem on an appropriate Hilbert space
and verifying that the evolution vector field generates a strongly
continuous semigroup. Possible generalizations to other formulations
of the linearized Einstein equations and future work are discussed in
the conclusions, Sect. \ref{Sect:Conclusions}. Appendix \ref{App:Fred}
summarizes standard results from Fredholm theory used in the
discussion of the gauge condition for the lapse, and Appendix
\ref{App:Proofs} contains technical proofs of some of the statements
made in Sect. \ref{Sect:WP}.

\section{Main ideas}
\label{Sect:MainIdeas}

In this section we sketch some ideas for obtaining a priori estimates
in the IBVP of Einstein's field equations. We start in
Sect. \ref{SubSect:EM} by reviewing the electromagnetic toy model
problem analyzed in \cite{oRoS05} and showing that the gauge condition
used in that work can be obtained by variation of a simple
functional. In Sect. \ref{SubSect:EFEQ} we decompose Einstein's field
equations into evolution and constraint equations. Next, in
Sect. \ref{SubSect:CV}, the constraint propagation system describing
the evolution of the constraint fields is derived and a boundary
condition which yields estimates for these fields is constructed. In
Sect. \ref{SubSect:WC} we discuss how estimates for the curvature
fields can be obtained. The curvature tensor can be decomposed into
the Ricci tensor and the Weyl tensor. The Ricci tensor can be
estimated through Einstein's equations provided suitable estimates on
the matter fields are available. Therefore, it remains to estimate the
Weyl tensor. In order to do so, we analyze its propagation and explain
the idea for constructing boundary conditions and deriving estimates
for the Weyl tensor following the lines of Ref. \cite{hFgN99}. Next,
in Sect. \ref{SubSect:FSFF}, the first and second fundamental forms
are estimated. For this, an appropriate gauge condition for the lapse
is needed. Such a condition will be derived by introducing a
functional in analogy with the electromagnetic case. Finally, the
resulting IBVP for Einstein's vacuum field equations is summarized in
Sect. \ref{SubSect:IBVP}. The well posedness of this problem in the
weak field limit is proven in the subsequent sections.

In the following, we assume that $\Sigma$ is an open bounded subset of
$\Real^3$ with $C^\infty$ boundary $\partial\Sigma$. The spacetime
metric on $[0,T] \times \Sigma$ has signature $(-,+,+,+)$ and is
denoted by ${\bf g}$ and the unique metric-compatible, torsion-free
connection by $\nabla$. The induced three-metric and connection on
$\Sigma$ are denoted by ${\bf h}$ and $D$, respectively. The indices
$a$, $b$, $c$, $d$, $e$, $f$ denote spacetime indices running from $0$
to $3$ while $i$, $j$, $k$, $l$ denote spatial indices running from
$1$ to $3$. We shall use the notation $\hateq$ for equalities holding
on $\partial\Sigma$ or on $[0,T] \times \partial\Sigma$. The unit
outward normal one-form to $\partial\Sigma$ is denoted by $N_i$ and
$\gamma_i{}^j \hateq \delta_i{}^j - N_i N^j$ is the projector onto the
tangent space of $\partial\Sigma$.

\subsection{The electromagnetic case}
\label{SubSect:EM}

In \cite{oRoS05} Maxwell's equations are written as a first order
system for the magnetic potential $A_j$, the electric field $E_j$ and
the derivatives of the magnetic potential, $W_{ij} = D_i A_j$. In the
source-free case, this system is
\begin{eqnarray}
\partial_t A_j &=& E_j + D_j\phi, 
\label{Eq:Af}\\ 
\partial_t E_j &=& D^i W_{ij} - (1+\sigma) D^i W_{ji} 
 + \sigma h^{kl} D_j W_{kl}\; ,
\label{Eq:Ef}\\
\partial_t W_{ij} &=& D_i E_j + \frac{\tau}{2}\, h_{ij} D^k E_k 
 + D_i D_j\phi,
\label{Eq:Wf}
\end{eqnarray}
subject to the constraints $C \equiv D^i E_i = 0$ and $C_{ij} \equiv
W_{ij} - D_i A_j = 0$. Here, $\phi$ denotes the electrostatic
potential, $h_{ij}$ the components of the Euclidean metric and
$\sigma$ and $\tau$ are nonvanishing constants having the same sign.

A well posed IBVP for this system is obtained in the following way:
First, the constraint propagation system describing the time evolution
of the constraint fields $C$ and $C_{ij}$ is analyzed. It is given by
\begin{eqnarray}
\partial_t C_{ij} &=& \frac{\tau}{2}\, h_{ij} C,
\label{Eq:Cijf}\\
\partial_t C &=& -\sigma\, D^i C_i\; ,
\label{Eq:Cf}\\ 
\partial_t C_i &=& -\tau\, D_i C,
\label{Eq:Ckf}
\end{eqnarray}
where $C_i \equiv D^j C_{ij} - h^{kl} D_i C_{kl}$. By imposing, for
instance, the boundary condition $N^i C_i \hateq 0$ one obtains an
$L^2$ estimate for the constraint fields. Next, using this, the
physical energy of the system and the radiative-type boundary
condition
\begin{equation}
\gamma_i^{\;\; j} E_j - N^j (W_{ij} - W_{ji}) \hateq g_i \; ,
\label{Eq:EMSommerfeld}
\end{equation}
where $g_i$ is a given vector field on the boundary satisfying $N^i
g_i \hateq 0$, one obtains $L^2$ bounds for the electric and magnetic
fields, $E_i$ and $W_{[ij]}$, respectively. The remaining difficulty
is to estimate the symmetric, gauge-dependent, part of $W_{ij}$. It
turns out that the simplest gauge choice, the temporal gauge $\phi =
0$, does not lead to such an estimate. This can be seen by means of
the following family of electrostatic solutions \cite{oRoS05}
\begin{eqnarray}
A_j &=& t\, D_j f, \nonumber\\
E_j &=& D_j f, \label{Eq:SolNotL2}\\
W_{ij} &=& t D_i D_j f \nonumber, 
\end{eqnarray}
where $f$ is a smooth, time-independent, harmonic function. This
family solves the evolution and constraint equations and the
radiative-type boundary condition with boundary data $g_i \hateq
\gamma_i{}^j D_j f$. A problem is the time-dependence of the magnetic
potential which implies that for $t > 0$ the solution depends on
second derivatives of $f$ while the initial and boundary data depends
only on first derivatives of $f$. This violates the expected energy
estimate in $L^2$.

For this reason, in \cite{oRoS05}, a different gauge condition was
introduced which can be obtained by minimizing the functional
\begin{displaymath}
I_{EM}[\phi] = \frac{1}{2}\int_{\Sigma} h^{ij} (\partial_t A_i)
(\partial_t A_j) \sqrt{h}\, d^3 x 
 = \frac{1}{2} \int_{\Sigma} h^{ij} (E_i + D_i\phi)(E_j + D_j\phi)
\sqrt{h}\, d^3 x
\end{displaymath}
over the space $H^1(\Sigma)$ consisting of square-integrable functions
$\phi$ on $\Sigma$ which have square-integrable first order spatial
derivatives. Therefore, this gauge condition minimizes the $L^2$ norm
of the time-deformation of the magnetic potential over all possible
electrostatic potentials $\phi$ for which this norm is defined. Here
and in the following, $\sqrt{h}$ and $h^{ij}$ denote, respectively,
the determinant and the inverse of the three-metric. In the above
example, where $E_j = D_j f$, the minimum of $I_{EM}$ is zero and
yields $\phi = -f$ which in turn implies that $A_j$ is
time-independent. Therefore, this gauge condition precludes the above
counterexample. Furthermore, it was shown in \cite{oRoS05} to yield a
well posed IBVP. A straightforward generalization of $I_{EM}$ is the
family of functionals
\begin{displaymath}
I_{EM,\mu}[\phi] = \frac{1}{2}\int_{\Sigma} \left[
 \mu^2\phi^2 + h^{ij}(\partial_t A_i)(\partial_t A_j) \right] \sqrt{h}\, d^3 x
\end{displaymath}
parametrized by a nonnegative constant $\mu$. The extra term,
$\mu^2\phi^2$, in the integrand plays the role of a penalty term which
prevents $|\phi|$ from becoming ``too large'' and might be useful for
numerical simulations. Variation of $I_{EM,\mu}$ with respect to
$\phi$ yields the elliptic boundary value problem
\begin{eqnarray}
 (-\mu^2 + D^i D_i)\phi + D^i E_i = 0 && \hbox{on $\Sigma$,}
\label{Eq:EMEllGaugeEq}\\
 N^i(D_i\phi + E_i) \hateq 0 && \hbox{on $\partial\Sigma$.}
\label{Eq:EMEllGaugeBC}
\end{eqnarray}
Notice that for $\mu > 0$ the solution of this boundary value problem
is unique while for $\mu=0$ the solution $\phi$ is only unique up to
an additive constant which might cause problems in an elliptic solver.
The important point here is the boundary condition
(\ref{Eq:EMEllGaugeBC}) which allows to set the normal component of
the magnetic potential to zero at the boundary: $N^j A_j \hateq
0$. This boundary condition in turn allows to estimate the $L^2$ norm
of ${\bf A}$ and its first order spatial derivatives in terms of the
divergence and the curl of ${\bf A}$. The divergence of ${\bf A}$ can
be estimated by taking the divergence of Eq. (\ref{Eq:Af}) and using
Eq. (\ref{Eq:EMEllGaugeEq}). The curl of ${\bf A}$ is known since it
corresponds to the magnetic field $W_{[ij]}$ for which an estimate has
been obtained.

Using the methods of Ref. \cite{oRoS05} one can show that the IBVP
described by the evolution equations
(\ref{Eq:Af},\ref{Eq:Ef},\ref{Eq:Wf}) for $\sigma\,\tau > 0$, the
boundary conditions (\ref{Eq:EMSommerfeld}) and $N^i C_i \hateq 0$ and
the gauge condition (\ref{Eq:EMEllGaugeEq},\ref{Eq:EMEllGaugeBC}) is
well posed. The interesting point is that one still obtains a well
posed problem in the limiting case $\sigma=\tau=0$ where the evolution
equations are weakly hyperbolic \cite{gNoS-inprep}. In this case, the
constraints propagate tangentially to the boundary, and the boundary
condition $N^i C_i \hateq 0$ has to be dropped.

\subsection{Einstein's field equations}
\label{SubSect:EFEQ}

Next, we discuss how the ideas above can be applied to the IBVP of
General Relativity. In the remaining part of this section, we consider
the full nonlinear Einstein equations with matter fields and draw
particular attention to the propagation of the constraint fields,
the propagation of the Weyl tensor and the gauge conditions. An IBVP
for the nonlinear case is mentioned at the end of this section. In the
next two sections, we prove that this problem is well posed in the
weak field regime.

We start with Einstein's field equations in the $3+1$ formulation of
Ref. \cite{jY79}. This is done for simplicity; as discussed in the
conclusions, other $3+1$ formulations might serve as a starting
point. Assume spacetime $(M,g)$ is globally hyperbolic, i.e. there
exists a globally defined time function $t: M \to I$ such that $M$ is
foliated by spacelike hypersurfaces $\Sigma_\tau = \{ p\in M : t(p) =
\tau \}$. Let $n_a = -\alpha \nabla_a t$ be the future-pointing unit
normal to these slices (we choose the time orientation such that the
lapse function, $\alpha$, is strictly positive). The three-metric
$h_{ab}$ and extrinsic curvature $k_{ab}$ (first and second
fundamental forms) are defined by
\begin{eqnarray}
h_{ab} &=& g_{ab} + n_a n_b\; ,\\
k_{ab} &=& -\nabla_a n_b - n_a a_b\; ,
\end{eqnarray}
respectively, where $a_b = n^a\nabla_a n_b = D_b(\log\alpha)$ is the
acceleration field. The $3+1$ form of Einstein's equations yield the
evolution equations
\begin{eqnarray}
\lie_n h_{ab} &=& -2k_{ab}\; ,
\label{Eq:EinsteinEvolutionh}\\
\lie_n k_{ab} &=& R_{ab} - \frac{1}{\alpha} D_a D_b\alpha + k k_{ab} -
 2k_a{}^d k_{bd} - \kappa s_{ab}\; ,
\label{Eq:EinsteinEvolutionk}
\end{eqnarray}
and the constraints
\begin{eqnarray}
H &\equiv& \frac{1}{2}
 \left( h^{ab} R_{ab} + k^2 - k^{ab} k_{ab} \right) - \kappa\rho = 0,
\label{Eq:EinsteinConstraintH}\\
M_a &\equiv& D_a k - D^b k_{ab} + \kappa j_a = 0,
\label{Eq:EinsteinConstraintM}
\end{eqnarray}
where here and in the following, $D_a$ and $R_{ab}$ denote,
respectively, the covariant derivative and the Ricci tensor compatible
with the three-metric $h_{ab}$, $k = h^{ab} k_{ab}$ is the trace of
the extrinsic curvature, $\kappa = 8\pi G$ where $G$ is the
gravitational constant, and where in terms of the stress-energy tensor
$T_{ab}$ the source terms $\rho$, $j_a$ and $s_{ab}$ are given by
\begin{eqnarray}
\rho &=& n^a n^b T_{ab}\; ,\nonumber\\
j_a &=& -h_a{}^c n^d T_{cd}\; ,\nonumber\\
s_{ab} &=& (h_a{}^c h_b{}^d - \frac{1}{2} h_{ab} g^{cd}) T_{cd} \; .
\nonumber
\end{eqnarray}
For consistency with the Bianchi identities, the stress-energy tensor
must be divergence-free which translates into the conditions
\begin{eqnarray}
&& \lie_n\rho + \frac{1}{\alpha^2} D^a\left( \alpha^2 j_a \right)
  - 2k\rho - (k^{ab} - k h^{ab}) s_{ab} = 0,
\label{Eq:Continuity}\\
&& \lie_n j_a + \frac{1}{\alpha^2} D_a\left( \alpha^2\rho \right)
  - k j_a + \frac{1}{\alpha}D^b
   \left( \alpha s_{ab} - \alpha h_{ab} h^{cd} s_{cd} \right) = 0.
\label{Eq:Stress}
\end{eqnarray}
A well-known property of the evolution system
(\ref{Eq:EinsteinEvolutionh},\ref{Eq:EinsteinEvolutionk}) is that the
Ricci operator, $h_{ab} \mapsto R_{ab}$, is not elliptic. In
particular, this implies that simple algebraic gauge choices yield a
weakly hyperbolic system for which there are no general well posedness
results and standard discretizations lead to unstable numerical
schemes \cite{KL-Book,GKO-Book,gCjPoSmT02a}. A possible remedy to this
problem is to adopt different gauge conditions, like the full harmonic
gauge \cite{DeDonder-Book}, in which the evolution equations become
manifestly hyperbolic.

\subsection{Estimates for the constraint fields}
\label{SubSect:CV}

Independent on what gauge is chosen, the evolution system
(\ref{Eq:EinsteinEvolutionh},\ref{Eq:EinsteinEvolutionk}) has the
property that it induces ``nice'' evolution equations for the
constraint fields: As a consequence of the (twice contracted) Bianchi
identity, the evolution equations
(\ref{Eq:EinsteinEvolutionh},\ref{Eq:EinsteinEvolutionk}) and the
consistency conditions (\ref{Eq:Continuity},\ref{Eq:Stress}) one finds
that $H$ and $M_a$ propagate according to the simple symmetric
hyperbolic system
\begin{eqnarray}
\lie_n H &=& \frac{1}{\alpha^2} D^a\left( \alpha^2 M_a \right) + 2k H,
\label{Eq:EvolH}\\
\lie_n M_a &=& \frac{1}{\alpha^2} D_a\left( \alpha^2 H \right) + k M_a \; ,
\label{Eq:EvolM}
\end{eqnarray}
which closely resembles a wave equation written in first order form.
By specifying suitable boundary conditions for this system, estimates
for the constraint fields $H$ and $M_a$ can be obtained. In the
following, we shall assume that the timelike boundary ${\cal T}$ is
such that $n^a\nu_a \geq 0$ everywhere on ${\cal T}$, where $\nu_a$
denotes the outward unit normal\footnote{Different boundary conditions
have to be specified for the case where $n^a\nu_a < 0$ everywhere on
${\cal T}$. We do not analyze this case here.} to ${\cal T}$. As can
be seen by a simple energy argument (cf. Lemma \ref{Lem:WPCPS} below),
the boundary condition
\begin{equation}
\nu^a M_a \hateq 0
\label{Eq:CPBC}
\end{equation}
guarantees that $H=0$ and $M_a=0$ everywhere on $[0,T] \times \Sigma$
provided $H=0$ and $M_a=0$ on the initial surface $\{ 0 \} \times
\Sigma$. Furthermore, the boundary condition (\ref{Eq:CPBC}) yields an
$L^2$ estimate for the constraint fields $H$ and $M_a$. As we will see
shortly, it turns out that we need more regularity and also need an
$L^2$ estimate for the first spatial derivatives of $H$ and $M_a$. It
will be shown in the linearized case that this can be achieved using
the boundary condition (\ref{Eq:CPBC}).

Before we proceed, we make two comments. First, we stress that the
evolution equations derived by Arnowitt-Deser-Misner \cite{ADM62}
differ from the one discussed here by the addition of the Hamiltonian
constraint to the evolution equations for $k_{ab}$. As a consequence,
in the former case, the resulting constraint propagation system is
only weakly hyperbolic which makes it much more difficult to control
the constraint fields. (See Ref. \cite{sF97} for the details on
this important difference.) Second, we would like to mention that if
$n^a\nu_a \hateq 0$ and in the absence of matter fields, the boundary
condition (\ref{Eq:CPBC}) corresponds to one of the four Einstein
boundary conditions $\nu^a G_{ab} \hateq 0$, where $G_{ab}$ denotes
the Einstein tensor, proposed in Ref. \cite{sFrG03b}. (In vacuum
Eq. (\ref{Eq:CPBC}) is equivalent to $\nu^a n^b G_{ab} \hateq 0$.)
However, the imposition of all four Einstein boundary conditions can
lead to difficulties. For example, imposing the two conditions $\nu^a
n^b G_{ab}\hateq 0$ and $\nu^a \nu^b G_{ab} \hateq 0$ overdetermines
the constraint propagation system since by virtue of the evolution
equations the second condition is equivalent to setting $H \hateq 0$.
Therefore, if the initial data violates the constraints (as is the
case in most numerical simulations), a solution to the IBVP might not
even exist.

\subsection{Estimates for the Weyl curvature}
\label{SubSect:WC}

The next step is to obtain estimates for the Weyl curvature tensor
$C_{abcd}$. In terms of the electric and magnetic part of the Weyl
tensor, defined by
\begin{eqnarray}
E_{bd} &\equiv& C_{abcd} n^a n^c, \\
B_{bd} &\equiv& \frac{1}{2} n^a C_{abef}\epsilon^{ef}{}_d\; ,
\end{eqnarray}
where $\epsilon_{bcd} \equiv n^a\epsilon_{abcd}$ denotes the natural
volume element on $(\Sigma_t, h_{ab})$, the Bianchi identities yield
\begin{eqnarray}
\lie_n E_{ab} &=& -\epsilon_{cd(a} (D^c + 2a^c) B^d{}_{b)}
\nonumber\\
 &-& 3k_{(a}{}^d E_{b)d} - 2k^d{}_{(a} E_{b)d} + 2k E_{ab} 
  + h_{ab} k^{cd} E_{cd} + P_{ab}\; ,
\label{Eq:EvolWeylE}\\
\lie_n B_{ab} &=& +\epsilon_{cd(a} (D^c + 2a^c) E^d{}_{b)}
\nonumber\\
 &-& 3k_{(a}{}^d B_{b)d} - 2k^d{}_{(a} B_{b)d} + 2k B_{ab} 
  + h_{ab} k^{cd} B_{cd} + Q_{ab}\; ,
\label{Eq:EvolWeylB}
\end{eqnarray}
where
\begin{eqnarray}
P_{ab} &=& \frac{1}{2}\left[ D_{(a} M_{b)} + 2 a_{(a} M_{b)} + 2k_{ab} H
\right. \nonumber\\
&& \qquad \left. - \kappa\left( \lie_n s_{ab} + k_{(a}{}^d s_{b)d} 
 + k_{ab} h^{cd} s_{cd} \right) \right]^{TF}
\label{Eq:PabDef}\\
Q_{ab} &=& -\frac{1}{2}\left[ \epsilon_{cd(a} k^c{}_{b)} M^d
                      + \kappa\epsilon_{cd(a} D^c s^d{}_{b)} \right],
\label{Eq:QabDef}
\end{eqnarray}
and the constraints on $E$ and $B$
\begin{eqnarray}
D^b E_{ab} + 2k^{cd}\epsilon^b{}_{da} B_{cb} &=& P_a\; ,\\
D^b B_{ab} - 2k^{cd}\epsilon^b{}_{da} E_{cb} &=& Q_a\; ,
\end{eqnarray}
where
\begin{eqnarray}
P_a &=& \frac{1}{3} D_a H 
 + \frac{1}{2}\left[ k_a{}^b M_b - k M_a
 - \kappa\left( D^b s_{ab} - \frac{1}{3} h^{cd} D_a s_{cd} \right) \right],
\label{Eq:PaDef}\\
Q_a &=& -\frac{1}{2}\epsilon_a{}^{cd}\left[ D_c M_d 
 + \kappa k_c{}^b s_{db} \right].
\label{Eq:QaDef}
\end{eqnarray}
In the above expressions for $P_a$, $Q_a$, $P_{ab}$ and $Q_{ab}$ we
have used the constraint propagation equations
(\ref{Eq:EvolH},\ref{Eq:EvolM}) in order to eliminate the Lie
derivatives of the constraint fields $H$ and $M_a$ with respect to
$n^a$.

Eqs. (\ref{Eq:EvolWeylE},\ref{Eq:EvolWeylB}) constitute a symmetric
hyperbolic system for the fields $E$ and $B$ which bears a striking
resemblance to Maxwell's equations. Notice that the source terms
$P_{ab}$ and $Q_{ab}$ depend on the constraint fields $H$ and $M_a$
and their first spatial derivatives as well as on first derivatives of
the matter terms $s_{ab}$. This is the reason why more regularity is
needed for the constraint fields (and the matter fields). A well
posed IBVP for the subsystem (\ref{Eq:EvolWeylE},\ref{Eq:EvolWeylB})
can be constructed using the method in Ref. \cite{hFgN99} where
suitable combinations of the constraint equations are added to the
evolution equations (\ref{Eq:EvolWeylE},\ref{Eq:EvolWeylB}) (see also
Sect. \ref{SubSect:WeylPS}). Boundary conditions for the resulting
linear evolution system for $E$ and $B$ which lead to a well posed
problem in $L^2$ can then be specified in the following form
\cite{hFgN99}: Set $N^a = h^{ai} N_i$ and choose for each $t\in [0,T]$
two unit vector fields $e_2^a$ and $e_3^a$ on $\{ t \} \times
\partial\Sigma$ which are mutually orthogonal, such that $N^a$,
$e_2^a$, $e_3^a$ form an orthonormal set. Define $m^a = 2^{-1/2}(e_2^a
+ i e_3^a)$ and introduce the complex Newman-Penrose scalars
\begin{eqnarray}
\Psi_0 &\hateq& \left[ E_{ab} - \epsilon_{cda} N^c B^d{}_b \right] 
m^a m^b,
\label{Eq:WeylScalar0}\\
\Psi_4 &\hateq& \left[ E_{ab} + \epsilon_{cda} N^c B^d{}_b \right]
\bar{m}^a \bar{m}^b.
\label{Eq:WeylScalar4}
\end{eqnarray}
Next, let $c$ be a smooth complex-valued function on $[0,T] \times
\partial\Sigma$ with magnitude smaller or equal than one, and let $q$
be a smooth complex-valued function on $[0,T] \times \partial\Sigma$.
Boundary conditions can then be specified in the form
\cite{hFgN99,oSmT05}
\begin{equation}
\Psi_0 \hateq c\bar{\Psi}_4 + q,
\label{Eq:PhysBC}
\end{equation}
where an overbar denotes complex conjugation. The appearance of the
complex conjugation in Eq. (\ref{Eq:PhysBC}) can be understood as
follows: Under a rotation $m \mapsto e^{i\varphi} m$ of $m$, $\Psi_0
\mapsto e^{2i\varphi}\Psi_0$ while $\Psi_4 \mapsto
e^{-2i\varphi}\Psi_4$. Therefore, if $q \mapsto e^{2i\varphi} q$, the
boundary condition (\ref{Eq:PhysBC}) is invariant with respect to such
rotations. Asymptotically, $\Psi_0$ and $\Psi_4$ represent,
respectively, the amount of in- and outgoing radiation. Therefore, the
function $c$ can be interpreted as a reflection coefficient.

\subsection{Estimates for the first and second fundamental forms}
\label{SubSect:FSFF}

So far, we have obtained estimates for the constraint fields and
the curvature fields $E_{ab}$ and $B_{ab}$. It remains to estimate the
three-metric $h_{ab}$ and the extrinsic curvature $k_{ab}$. For this,
we first notice that the electric and magnetic components of the Weyl
tensor can be expressed as
\begin{eqnarray}
E_{ab} &=& \frac{1}{2}\left[ \lie_n k_{ab} + R_{ab}
 + \frac{1}{\alpha} D_a D_b \alpha + k k_{ab} \right]^{TF}, 
\label{Eq:EDef}\\
B_{ab} &=& \epsilon_{cd(a} D^c k^d{}_{b)}\; ,
\label{Eq:BDef}
\end{eqnarray}
where $[ ... ]^{TF}$ denotes the trace-free part with respect to
$h_{ab}$. This allows us to rewrite the evolution system
(\ref{Eq:EinsteinEvolutionh},\ref{Eq:EinsteinEvolutionk}) as
\begin{eqnarray}
\lie_n h_{ab} &=& -2k_{ab}\; ,
\label{Eq:EinsteinEvolutionhbis}\\
\lie_n k_{ab} &=& E_{ab} - \frac{1}{\alpha} D_a D_b\alpha
 - k_a{}^c k_{cb} + \frac{2}{3} h_{ab} H
\nonumber\\
&-& \frac{\kappa}{2} s_{ab} 
 + \frac{\kappa}{6} h_{ab} \left( 4\rho - h^{cd} s_{cd} \right).
\label{Eq:EinsteinEvolutionkbis}
\end{eqnarray}
Since estimates for $E_{ab}$ and $H$ have already been obtained, and
provided suitable estimates for the matter terms $\rho$ and $s_{ab}$
are available, one should therefore be able to integrate this system
and obtain estimates for $h_{ab}$ and $k_{ab}$. The question here is
what precise estimates we need. Since one would like to be able to
estimate the Christoffel symbols, we need to control at least the
$L^2$ norm of the three-metric and its first order spatial
derivatives. Such an estimate can be obtained for fixed shift by
integrating Eq. (\ref{Eq:EinsteinEvolutionhbis}) provided a similar
estimate for $k_{ab}$ is available. As in the electromagnetic case,
one does not expect such an estimate to exist for any gauge condition
for the lapse. Comparing Eq. (\ref{Eq:Af}) with
Eq. (\ref{Eq:EinsteinEvolutionkbis}) and making the analogy $A_i
\leftrightarrow k_{ab}$, $E_i \leftrightarrow E_{ab}$, $\phi
\leftrightarrow \alpha$ between the electromagnetic case and General
Relativity, the functional $I_{EM}$ suggests the consideration of the
following functional over the space $H^2(\Sigma)$ consisting of
square-integrable functions $\alpha$ on $\Sigma$ with
square-integrable first and second order spatial derivatives,
\begin{eqnarray}
I_{GR}[\alpha] &=& \frac{1}{2}\int_{\Sigma}
   h^{ac} h^{bd} \left( \frac{\partial_t k_{ab} }{\alpha} \right)
                 \left( \frac{\partial_t k_{cd} }{\alpha} \right)
\sqrt{h}\, d^3 x
\nonumber\\
 &=& \frac{1}{2}\int_{\Sigma} h^{ac} h^{bd} 
   \left( \frac{1}{\alpha} D_a D_b\alpha 
        - \frac{1}{\alpha}\lie_\beta k_{ab} - F_{ab} \right)
\nonumber\\
&& \qquad\qquad\qquad
   \left( \frac{1}{\alpha} D_c D_d\alpha 
        - \frac{1}{\alpha}\lie_\beta k_{cd} - F_{cd} \right)
\sqrt{h}\, d^3 x,
\nonumber
\end{eqnarray}
where $F_{ab} = R_{ab} + k k_{ab} - 2k_a{}^d k_{bd} - \kappa s_{ab}$
and $\beta^a$ is the shift vector field. The reason for introducing
the factor $1/\alpha$ in front of $\partial_t k_{ab}$ is that without
this factor, the minimum of $I_{GR}$ is always zero and is attained
for $\alpha=0$ when $\lie_\beta k_{ab}$ vanishes\footnote{This problem
could be avoided by requiring an inhomogeneous boundary condition for
the lapse; however, for our key estimate in Lemma
\ref{Lem:KeyEstimate} it turns out that we need to vary $I_{GR}$ over
the space of all $\alpha\in H^2(\Sigma)$.}. Notice that minima need
not be unique: For example, if $\lie_\beta k_{ab} = 0$, the functional
is invariant with respect to the rescaling $\alpha \mapsto
\lambda\alpha$, where $\lambda$ is a nonvanishing constant. This
rescaling corresponds to the coordinate reparametrization $t \mapsto
\lambda^{-1} t$. Similarly to the electromagnetic case, the functional
$I_{GR}$ can be generalized to
\begin{displaymath}
I_{GR,\mu}[\alpha] = \frac{1}{2}\int_{\Sigma}
\left[ \mu^2(\alpha-\alpha_0)^2 
 + h^{ac} h^{bd} \left( \frac{\partial_t k_{ab} }{\alpha} \right)
                 \left( \frac{\partial_t k_{cd} }{\alpha} \right)
\right] \sqrt{h}\, d^3 x
\end{displaymath}
where here, $\mu \geq 0$ and $\alpha_0$ is a given, strictly positive
function on spacetime. The penalty term $\mu^2(\alpha-\alpha_0)^2$
might be useful in order to guarantee the uniqueness of minima and to
ensure that the lapse is positive. Below, we analyze the well
posedness of the variational principle associated to $I_{GR,\mu}$ in
the weak field limit and show that the resulting condition for the
lapse leads to an $L^2$ estimate for $k_{ab}$ and its first order
spatial derivatives (see Lemma \ref{Lem:KeyEstimate} below).

A similar functional was recently used in \cite{sD04} in order to
construct a new geometric invariant measuring the amount of radiation
contained in a data set and was also considered in \cite{cBlLcP05}.
Functionals very similar to $I_{GR}$ also arise in the theory of thin
elastic plates (see, for instance, \cite{Ciarlet-Book}).

\subsection{The IBVP for Einstein's vacuum field equations}
\label{SubSect:IBVP}

With the above considerations we may formulate the following IBVP for
the Einstein vacuum field equations. The generalization to matter
fields depends on the precise model for the matter. For this reason,
from now on, we set $T_{ab} = 0$.
\begin{enumerate}
\item Specify initial data $(h_{ij},k_{ij})$ on a three-dimensional
Riemannian compact manifold with smooth boundary $\partial\Sigma$
satisfying the Hamiltonian and momentum constraint equations
(\ref{Eq:EinsteinConstraintH},\ref{Eq:EinsteinConstraintM}).
\item Specify boundary data $q(t)$ on $[0,T] \times \partial\Sigma$.
The initial and boundary data must satisfy certain compatibility
conditions on the intersection $\{ 0 \} \times \partial\Sigma$ of the
initial surface with the timelike boundary (see, for example
\cite{pS96a,pS96b}).
\item Choose the shift as an a priori specified vector field $\beta^i$
on $[0,T] \times \Sigma$ such that $\beta^i N_i \leq 0$ everywhere on
${\cal T}$. Notice that this inequality is independent of the (yet
unknown) metric.
\item Find a curve $t \mapsto (h_{ij},k_{ij})(t)$, $0 \leq t \leq T$,
solution of the evolution equations
(\ref{Eq:EinsteinEvolutionh},\ref{Eq:EinsteinEvolutionk}), where in
Eq. (\ref{Eq:EinsteinEvolutionk}) the lapse is given by minimizing the
functional $I_{GR,\mu}$, such that
\begin{equation}
(h_{ij},k_{ij})(0) =  (h_{ij},k_{ij})
\end{equation}
and such that the boundary conditions
\begin{equation}
N^i M_i(t) \hateq 0, \qquad
\Psi_0(t) \hateq c\bar{\Psi}_4(t) + q(t),
\end{equation}
hold, where $M_i$, $\Psi_0$ and $\Psi_4$ are defined in terms of
$(h_{ij},k_{ij})$ according to Eq. (\ref{Eq:EinsteinConstraintM}) and
Eqs. (\ref{Eq:WeylScalar0},\ref{Eq:WeylScalar4},\ref{Eq:EDef},\ref{Eq:BDef}),
respectively.
\end{enumerate}
In the following, we analyze the well posedness of this problem under
simplifying assumptions. First, we only consider the weak field regime
in which all the equations are linearized about the flat spacetime
solution $(h_{ij},k_{ij}) = (\delta_{ij},0)$, $\alpha = 1$, $\beta^i =
0$. Second, we assume that the domain $\Sigma = (0,1) \times T^2$ is a
cube where the two opposite faces $y=const$ and $z=const$ are
identified with each other (although most of the results below are
valid for more general domains). The main result of this article is
the well posedness of the above stated IBVP under these simplifying
assumptions, see Theorem \ref{Thm:Main} below. Results for more
general cases will be discussed elsewhere \cite{gNoS-inprep}.

\section{Gauge condition for the lapse}
\label{Sect:LapseGauge}

In this section we analyze the well posedness of the variational
principle for $I_{GR,\mu}$. As indicated above, for simplicity, we
only consider here the case of linearization about flat spacetime. In
this case, it is natural to set $\alpha_0 = 1$, and we expand $\alpha
= 1 + f$ and ignore all terms in $I_{GR,\mu}$ which are cubic or
higher order in $f$ and the perturbed metric and extrinsic
curvature. This leads to the functional $I_\mu$ defined in
Eq. (\ref{Eq:Functional}) below where now $F_{ij}$ is equal to the
linearized Ricci tensor.

Before we proceed, we fix some notation: We denote by
$L^2(\Sigma,\Real)$, $L^2(\Sigma,V)$ and $L^2(\Sigma,S)$ the spaces of
square-integrable functions, vector fields and symmetric tensor
fields, respectively, on $\Sigma$. Similarly, $H^m(\Sigma,\Real)$,
$H^m(\Sigma,V)$, $H^m(\Sigma,S)$ denote the Sobolev spaces of order
$m$ of smooth functions, vector fields and symmetric tensor fields on
$\Sigma$ and $C^\infty(\bar{\Sigma},\Real)$,
$C^\infty(\bar{\Sigma},V)$, $C^\infty(\bar{\Sigma},S)$ the space of
functions, vector fields and symmetric tensor fields on $\bar{\Sigma}$
which are infinitely differentiable on $\Sigma$ and such that all
derivatives have a continuous extension on $\bar{\Sigma}$.

Define the bounded linear operator
\begin{displaymath}
W: H^2(\Sigma,\Real) \to L^2(\Sigma,S), \quad
   f \mapsto W_{ij} f = D_i D_j f.
\end{displaymath}
We consider for each $\mu \geq 0$ and ${\bf F}\in L^2(\Sigma,S)$ the
functional
\begin{eqnarray}
I_\mu &:& H^2(\Sigma,\Real) \to \Real
\nonumber\\
      && f \mapsto \frac{1}{2}\int_{\Sigma} \left[
   \mu^2 f^2 + \left( W^{ij} f - F^{ij} \right)
               \left( W_{ij} f - F_{ij} \right) \right] \sqrt{h}\, d^3 x.
\label{Eq:Functional}
\end{eqnarray}

\begin{theorem}[Minimum of $I_\mu$]
\label{Thm:Min}
The functional $I_\mu: H^2(\Sigma,\Real) \to \Real$ has the following
properties:
\begin{enumerate}
\item[(i)] There exists a global minimum $f\in H^2(\Sigma,\Real)$ of
$I_\mu$. For $\mu > 0$ this minimum is unique; for $\mu = 0$ the
minimum is unique up to the addition of an element in the
two-dimensional set
\begin{displaymath}
{\cal N} = \{ f = a_0 + a_1 x : a_0, a_1\in\Real \}.
\end{displaymath}
\item[(ii)] A global minimum $f\in H^2(\Sigma,\Real)$ of
$I_\mu$ satisfies
\begin{displaymath}
I_\mu[f] \leq \frac{1}{2} \| {\bf F} \|_{L^2(\Sigma,S)}^2.
\end{displaymath}
\item[(iii)] If $\mu=0$, a global minimum $f\in H^2(\Sigma,\Real)$
satisfies $({\bf F} - {\bf W}f, {\bf W}f)_{L^2(\Sigma,S)} = 0$,
i.e. ${\bf F} - {\bf W} f$ and ${\bf W}f$ are orthogonal.
\item[(iv)] If ${\bf F}\in C^\infty(\bar{\Sigma},S)$, all global
minima $f\in H^2(\Sigma,\Real)$ automatically lie in
$C^\infty(\bar{\Sigma},S)$ and satisfy the elliptic boundary value
problem
\begin{eqnarray}
\mu^2 f + W^{ij}(W_{ij} f - F_{ij}) = 0 && \hbox{on $\Sigma$,}
\nonumber\\
N^i N^j (W_{ij} f - F_{ij}) \hateq 0 && \hbox{on $\partial\Sigma$,}
\nonumber\\
N^i (D^j + {\cal D}^j)(W_{ij} f - F_{ij}) \hateq 0 && 
\hbox{on $\partial\Sigma$,}
\end{eqnarray}
where ${\cal D}$ denotes the covariant derivative on $\partial\Sigma$
induced by $D$.
\end{enumerate}
\end{theorem}

\proof
The proof is based on standard arguments from elliptic theory. In a
first step, we set $X \equiv H^2(\Sigma,\Real)$ and rewrite the
functional $I_\mu$ in the form
\begin{displaymath}
I_\mu[f] = Q_\mu(f,f) - 2J(f) + \frac{1}{2}\| {\bf F} \|_{L^2(\Sigma,S)}^2,
\end{displaymath}
for $f\in X$, where the bounded bilinear form $Q_\mu: X \times X \to \Real$ 
and the linear functional $J : X \to \Real$ are defined by
\begin{eqnarray}
Q_\mu(f,g) &=& \frac{1}{2}\left[ \mu^2(f,g)_{L^2(\Sigma,\Real)} 
 + ({\bf W}f,{\bf W}g)_{L^2(\Sigma,S)} \right],
\nonumber\\
J(g) &=& \frac{1}{2} ({\bf F},{\bf W}g)_{L^2(\Sigma,S)}\; ,
\nonumber
\end{eqnarray}
for $f,g\in X$. Since for all $f,g \in X$ and all $\varepsilon > 0$ we
have
\begin{equation}
\frac{I_\mu[f + \varepsilon g] - I_\mu[f]}{\varepsilon} 
 = 2(Q_\mu(f,g) - J(g)) + \varepsilon Q_\mu(g,g).
\label{Eq:IDiff}
\end{equation}
and since $Q_\mu(g,g)\geq 0$, it follows that $f\in X$ is a global
minimum of $I_\mu$ if and only if
\begin{equation}
Q_\mu(f,g) = J(g) \qquad\hbox{for all $g\in X$}.
\label{Eq:WeakSolution}
\end{equation}

(i) Therefore, we have to show that there exists $f\in X$ such that
Eq. (\ref{Eq:WeakSolution}) holds. This is done using the theorem in
Appendix \ref{App:Fred} which summarizes standard results from
Fredholm theory. In order to apply the theorem, we first notice that
\begin{displaymath}
Q_\mu(f,f) + \frac{1}{2} \| f \|^2_{H^1(\Sigma,\Real)} 
 \geq \frac{1}{2} \| f \|^2_{H^2(\Sigma,\Real)}
\end{displaymath}
for all $f\in X$. Next, set $Z = H^1(\Sigma,\Real)$ and denote by
$\iota: X \to Z$ the inclusion which is compact since $\Sigma$ is
bounded. Furthermore, we introduce the following linear bounded
operators
\begin{eqnarray}
M: X \to X^*, && f \mapsto Q_\mu(f,.),
\nonumber\\
L: X \to X^*, && f \mapsto Q_\mu(f,.) + \frac{1}{2}(f,.)_Z\; ,
\nonumber\\
R: Z \to X^*, && f \mapsto -\frac{1}{2}(f,.)_Z\; .
\nonumber
\end{eqnarray}
With this notation, $f\in X$ satisfies Eq. (\ref{Eq:WeakSolution}) if
and only if $M f = J$, and $M = L + K$ is the sum of the coercive
operator $L$ and the compact operator $K = R \circ \iota$. According
to Theorem \ref{Thm:Fred}, a solution to this equation exists if and
only if $J\in (\ker M^t)^\perp$, where $(M^t f)(g) = (M g)(f)$ for all
$f,g\in X$ and $(\ker M^t)^\perp$ denotes the annihilator of $\ker
M^t$. In our case, since $Q_\mu$ is symmetric,
\begin{displaymath}
(M^t f)(g) = Q_\mu(g,f) = Q_\mu(f,g) = (M f)(g)
\end{displaymath}
for all $f,g\in X$; hence $M^t = M$. If $\mu > 0$ we have $\ker M = \{
0 \}$ and so $(\ker M^t)^\perp = X^*$. Therefore, if $\mu > 0$, there
exists a unique global minimum $f\in X$. On the other hand, if $\mu =
0$,
\begin{displaymath}
\ker M = \{ f\in X : W_{ij}f = 0 \} = \{ a_0 + a_1 x : a_0, a_1\in\Real \}.
\end{displaymath}
Furthermore, by definition, $J\in (\ker M)^\perp$. Hence, the existence
of global minima follows from Theorem \ref{Thm:Fred}.

(ii) Setting $\varepsilon=1$, $g=-f$ in Eq. (\ref{Eq:IDiff}) we obtain
\begin{displaymath}
I_\mu[f] = I_\mu[0] - Q_\mu(f,f) \leq I_\mu[0] 
 = \frac{1}{2} \| {\bf F} \|_{L^2(\Sigma,S)}^2.
\end{displaymath}

(iii) This follows from Eq. (\ref{Eq:WeakSolution}) by setting $g=f$.

(iv) This can be seen by explicitly computing the solution and using
integration by parts.

\qed

Theorem \ref{Thm:Min} has the following physical interpretation:
Consider the linearized extrinsic curvature $k_{ij}$ and perform an
infinitesimal coordinate transformation $\delta x^a \mapsto \delta x^a
+ X^a$ parametrized by the vector field $(X^a) = (X^t,X^i)$. With
respect to this, the linearized extrinsic curvature transforms
according to
\begin{equation}
k_{ij} \mapsto k_{ij} - W_{ij} f,
\end{equation}
where $f = X^t$. Therefore, given ${\bf k}\in L^2(\Sigma,S)$, the
minimization principle described by $I_\mu$ provides a unique way of
gauge-fixing $k_{ij}^{(GF)} \equiv k_{ij} - W_{ij} f$ by choosing $f$
such that it minimizes
\begin{displaymath}
I_\mu[f] = \frac{1}{2} \int_\Sigma \left[ \mu^2 f^2 
 + h^{ik} h^{jl} (k^{(GF)}_{ij})(k^{(GF)}_{kl}) \right] \sqrt{h}\, d^3 x.
\end{displaymath}
According to Theorem \ref{Thm:Min}(ii), this defines for each $\mu
\geq 0$ a bounded linear map $P_\mu^{(GF)}: L^2(\Sigma,S) \to
L^2(\Sigma,S)$, ${\bf k} \mapsto {\bf k}^{(GF)}$ whose operator norm
is less or equal than one. In particular, we have a unique
decomposition
\begin{equation}
k_{ij} = k_{ij}^{(GF)} + W_{ij} f
\label{Eq:ExtCurvDecomp}
\end{equation}
for each ${\bf k}\in L^2(\Sigma,S)$, where ${\bf k}^{(GF)} \equiv
P_\mu^{(GF)} {\bf k}$. For $\mu=0$ this decomposition is orthogonal
and $P_0^{(GF)}$ is an orthogonal projector, i.e. $P_0^{(GF)} \circ
P_0^{(GF)} = P_0^{(GF)}$ and $(P_0^{(GF)})^* = P_0^{(GF)}$. If ${\bf
k}\in C^\infty(\bar{\Sigma},S)$, the same holds for ${\bf k}^{(GF)}$
and it satisfies
\begin{eqnarray}
\mu^2 f - W^{ij} k_{ij}^{(GF)} = 0 && \hbox{on $\Sigma$},\\
N^i N^j k^{(GF)}_{ij} \hateq 0 &&  \hbox{on $\partial\Sigma$,}\\
N^i (D^j + {\cal D}^j) k^{(GF)}_{ij} \hateq 0 && \hbox{on $\partial\Sigma$.}
\end{eqnarray}

One of the key properties of the gauge-fixing operator $P_\mu^{(GF)}$
that will be crucial in our well posedness proof below is the
following: Defining the curl and momentum operators by
\begin{eqnarray}
\curl &:& H^1(\Sigma,S) \to L^2(\Sigma,S) \nonumber\\
      && k_{ij} \mapsto (\curl {\bf k})_{ij} 
                        \equiv \varepsilon_{kl(i} D^k k^l{}_{j)}\; , 
\nonumber\\
\M &:& H^1(\Sigma,S) \to L^2(\Sigma,V) \nonumber\\
   && k_{ij} \mapsto (\M {\bf k})_i \equiv D_i k - D^j k_{ij}\; ,
\nonumber
\end{eqnarray}
the gauge-fixed linearized extrinsic curvature ${\bf k}^{(GF)} =
P_\mu^{(GF)}{\bf k}$ has the property that its $H^1$ norm is already
controlled by its $L^2$ norm and the $L^2$ norm\footnote{Notice that
$\curl {\bf k}^{(GF)} = \curl {\bf k}$ and $\M {\bf k}^{(GF)} = \M
{\bf k}$.} of $\curl {\bf k}^{(GF)}$ and $\M {\bf k}^{(GF)}$:

\begin{lemma}
\label{Lem:KeyEstimate}
Let $\mu \geq 0$. There exists a constant $C = C_\mu > 0$ such that
\begin{equation}
\| {\bf k} \|_{H^1(\Sigma,S)} \leq C\left[ \| {\bf k} \|_{L^2(\Sigma,S)} 
 + \| \curl {\bf k} \|_{L^2(\Sigma,S)} + \| \M {\bf k} \|_{L^2(\Sigma,V)} \right]
\label{Eq:KeyEstimate}
\end{equation}
for all ${\bf k} \in P_\mu^{(GF)} C^\infty(\bar{\Sigma},S)$.
\end{lemma}

\proof 
Set $(N^i) = (1,0,0)$ and denote by $A$, $B$, $C$, $D$, $E$, $F$
indices running over $y$ and $z$. Define the quantities
\begin{eqnarray}
Q &\equiv& N^i N^j (\curl {\bf k})_{ij} = \varepsilon^{AB} D_A k_{Bx}\; ,
\nonumber\\
R &\equiv& N^i (\M {\bf k})_i = D_x k^{B}{}_{B} - D^B k_{xB}\; ,
\nonumber\\
V^{(+)}_A &\equiv& \gamma_{Aj}\left[ -\frac{1}{2} (\M {\bf k})^j 
 + N_i\varepsilon^{ijk} (\curl {\bf k})_{kl} N^l \right]
 = D_x k_{xA} - D_A k_{xx}\; ,
\nonumber\\
V^{(-)}_A &\equiv& \gamma_{Aj}\left[ -\frac{1}{2} (\M {\bf k})^j 
 - N_i\varepsilon^{ijk} (\curl {\bf k})_{kl} N^l \right]
 = D^B k_{AB} - D_A k^B{}_B\; ,
\nonumber\\
q_{AB} &\equiv& -\gamma_{Ai} N_j \varepsilon^{ijk} (\curl {\bf k})_{kl} 
 \gamma^l{}_B = \left[ D_x k_{AB} - D_A k_{xB} \right]^{tf},
\nonumber
\end{eqnarray}
where $[ v_{AB} ]^{tf} \equiv v_{AB} -
\frac{1}{2}\delta_{AB}\delta^{CD} v_{CD}$ denotes the trace-free part
of a two-tensor $v_{AB}$. Using integration by parts, we first find
\begin{eqnarray}
&& \int\limits_\Sigma \left[ Q^2 + R^2 
 + \gamma^{AB}(V^{(+)}_A + D_A k)(V^{(+)}_B + D_B k) \right] d^3 x
\nonumber\\
 &=& \int\limits_\Sigma \left[ (D_x k^B{}_B)^2 + (D^A k^B{}_B)(D_A k^C{}_C)
 + (D^x k^{xB})(D_x k_{xB}) \right. \nonumber\\
 && \left. \qquad + (D^A k^{xB})(D_A k_{xB}) \right] d^3 x
 - 2\int\limits_{T^2} \left[ (k^B{}_B)(D^A k_{xA}) \right]_{x=0}^1 dy dz.
\label{Eq:Id1}
\end{eqnarray}
On the other hand, using the fact that $D^i D^j k_{ij} = \mu^2 f$, we
find
\begin{eqnarray}
&& 2\int\limits_\Sigma \left[ (D^i k)(\M {\bf k})_i - \mu^2 k f  \right] d^3 x
\nonumber\\
&=& 2\int\limits_\Sigma (D^i k)(D_i k) d^3 x 
 - 2\int\limits_{T^2} \left[ k D^j k_{xj} \right]_{x=0}^1 dy dz.
\label{Eq:Id2}
\end{eqnarray}
Adding Eqs. (\ref{Eq:Id1}) and (\ref{Eq:Id2}) together, and using the
boundary conditions $k_{xx} = 0$, $D^j k_{xj} + D^A k_{xA} = 0$ on
$\partial\Sigma$ we obtain
\begin{eqnarray}
&& \int\limits_\Sigma \left[ (D^i k^B{}_B)(D_i k^C{}_C) 
 + (D^i k^{xB})(D_i k_{xB}) + 2(D^i k)(D_i k) \right] d^3 x
\nonumber\\
 &=& \int\limits_\Sigma \left[ Q^2 + R^2 
 + \gamma^{AB}(V^{(+)}_A + D_A k)(V^{(+)}_B + D_B k) 
 + 2(D^i k)(\M {\bf k})_i - 2\mu^2 k f \right] d^3 x
\nonumber\\
 &\leq& \int\limits_\Sigma \left[ Q^2 + R^2 
 + (1 + K_1)\gamma^{AB} V^{(+)}_A V^{(+)}_B
 + K_2 (\M {\bf k})^i (\M {\bf k})_i \right] d^3 x
\nonumber\\
 &+& \left( \frac{1}{K_1} + \frac{1}{K_2} \right)
     \int\limits_\Sigma (D^i k)(D_i k) d^3 x
  + K_3 \| {\bf k} \|_{L^2(\Sigma,S)}^2,
\nonumber
\end{eqnarray}
where $K_1$, $K_2$ and $K_3$ are positive constants and where we have
used Theorem \ref{Thm:Min}(ii) in the last step in order to estimate
$\mu f$. By choosing $K_1$ and $K_2$ large enough such that $K_1^{-1}
+ K_2^{-1} < 2$ it follows that we can bound the $H^1$ norm of the
components $k_{xx}$, $k_{xA}$, $k^A{}_A$ in terms of the right-hand
side of (\ref{Eq:KeyEstimate}).

It remains to bound the $H^1$ norm of $\hat{k}_{AB} = k_{AB} -
\frac{1}{2} \delta_{AB} k^C{}_C$. For this, we first notice the
identity
\begin{displaymath}
\varepsilon_A{}^C \hat{k}_{CB} = -\hat{k}_{AC}\varepsilon^C{}_B
\end{displaymath}
which is valid for any symmetric, trace-less two-tensor
$\hat{k}_{AB}$. This implies
\begin{displaymath}
2D_{[A} \hat{k}_{B]C} = \varepsilon_{AB}\varepsilon^{EF}D_E\hat{k}_{FC}
 = -\varepsilon_{AB} D^E k_{ED}\varepsilon^D{}_C.
\end{displaymath}
Using this, we find
\begin{eqnarray}
&& \int\limits_\Sigma (D^i\hat{k}^{BC})(D_i\hat{k}_{BC}) d^3 x
\nonumber\\
 &=& \int\limits_\Sigma \left[ (D^x\hat{k}^{BC})(D_x\hat{k}_{BC}) 
 + (D^A\hat{k}^{BC})(D_B\hat{k}_{AC}) 
 + 2(D^A\hat{k}^{BC})(D_{[A}\hat{k}_{B]C}) \right]d^3 x
\nonumber\\
 &=& \int\limits_\Sigma \left[ (D^x\hat{k}^{BC})(D_x\hat{k}_{BC}) 
 + 2(D_B\hat{k}^{BC})(D^A\hat{k}_{AC}) \right]d^3 x
\nonumber\\
 &=& \int\limits_\Sigma \gamma^{AC}\gamma^{BD}
 \left[ q_{AB} + (D_A k_{xB})^{tf} \right]
 \left[ q_{CD} + (D_C k_{xD})^{tf} \right] d^3 x
\nonumber\\
 &+& \frac{1}{2}\int\limits_\Sigma \gamma^{AC}
 \left[ 2V^{(-)}_A + D_A k^B{}_B \right]
 \left[ 2V^{(-)}_C + D_C k^D{}_D \right] d^3 x.
\nonumber
\end{eqnarray}
where we have used integration by parts in the second step. This
proves the statement of the Lemma.
\qed

\section{Well posed IBVP for the linearized vacuum equations}
\label{Sect:WP}

In this section we present a well posed initial-boundary value
formulation for the linearized Einstein vacuum equations when
linearized about a flat spacetime. For simplicity, we also assume that
$\Sigma = (0,1) \times T^2$ although most of the results below hold
for the general case where $\Sigma \subset \Real^3$ is an open bounded
domain with $C^\infty$ boundary $\partial\Sigma$. In the following,
let
\begin{displaymath}
(N^j) = (1,0,0), \qquad
(m^j) = \frac{1}{\sqrt{2}}(0,1,i),
\end{displaymath}
such that $N^j$, $m^j$, $\bar{m}^j$ form a complex triad adapted to
the boundary. We assume that the linearized shift is exactly zero and
fix the gauge for the linearized lapse by requiring that $P^{(GF)}
{\bf k} = {\bf k}$ where $P^{(GF)} \equiv P_0^{(GF)}$ is the
gauge-fixing projection operator introduced in the previous
section. The linearized equations have the form
\begin{equation}
\frac{d}{dt} u = A u, 
\label{Eq:EvolutionEquation}
\end{equation}
where $u = ({\bf h},{\bf k})$ denotes the components of the linearized
three-metric and extrinsic curvature and
\begin{equation}
A\left( \begin{array}{c} {\bf h} \\ {\bf k} \end{array} \right)
 = \left( \begin{array}{c} -2{\bf k} \\ 
       P^{(GF)} \Ric {\bf h} \end{array} \right),
\label{Eq:EvolutionVectorField}
\end{equation}
where $\Ric$ is the linearized Ricci operator. The idea is to
represent $A$ on an appropriate Hilbert space $H$ such that $A:
D_H\subset H \to H$ generates a strongly continuous
semigroup\footnote{That is, $P$ is a strongly continuous map from
$[0,\infty)$ to ${\cal L}(H)$ satisfying $P(0) = \id_H$ and the
semigroup property $P(t+s) = P(t)\circ P(s)$ for all $t,s\geq 0$.}
$P(t) = \exp(t A)$ on $H$. The unique solution of the abstract Cauchy
problem
\begin{eqnarray}
& \dfrac{d}{dt}u = Au, & t > 0,
\nonumber\\
& u(t=0) = u_0\; ,    & u_0\in D_H
\label{Eq:CauchyProblem}
\end{eqnarray}
is then given by $u(t)=P(t)u_0$, $t \geq 0$. Furthermore, the
semigroup properties imply the existence of constants $a\geq 1$, $b\in
\Real$ such that $\| u(t) \| \leq a\exp(b t) \| u_0 \|$ for all $t\geq
0$ and all $u_0\in H$. In particular, the problem is well posed.

Therefore, our task is to find a Hilbert space $H$ and a domain $D_H
\subset H$ such that $A: D_H \subset H \to H$ generates a strongly
continuous semigroup. There are well-known sufficient and necessary
conditions for this to be the case, see for instance \cite{Pazy-Book}
and \cite{hB05} and references therein for generalizations to
quasilinear operators. In our case, the Hilbert space $H$ is motivated
from our general considerations in Sect. \ref{Sect:MainIdeas} and we
will first define $A$ on the space of smooth tensor fields $({\bf
h},{\bf k})$ satisfying the boundary conditions (the space $D_0$
below). Then, it will be shown that $A$ is
quasi-dissipative\footnote{$A$ is called dissipative if $(u,Au) \leq
0$ for all $u\in D(A)$ and quasi-dissipative if there exists a
constant $b\in\Real$ such that $A-b$ is dissipative.} and that it can
be extended (by taking its closure in $H$) in such a way that the
extension generates a strongly continuous semigroup on $H$. The
following theorem gives the general structure of our Hilbert space
$H$.

\begin{theorem}[Linear constrained evolution systems with curvature map]
\label{Thm:LinConstrainedEvolutionCurvMap}
Let $X,Y,Z$ be real Hilbert spaces, and let
\begin{eqnarray}
A: D(A) \subset X \to X  
&& \hbox{(main evolution vector field)},\\
\hat{A}: D(\hat{A}) \subset Y \to Y 
&& \hbox{(constraint evolution vector field)},\\
B: D(B) \subset Z \to Z
&& \hbox{(curvature evolution vector field)},
\end{eqnarray}
be densely-defined linear operators on $X$, $Y$ and $Z$, respectively. 
Further, let $D_0\subset D(A)$ be a dense linear subspace of $X$ and
\begin{eqnarray}
S_0: D \subset X \to Y 
&& \hbox{(constraint map)},\\
T_0: D \subset X \to Z 
&& \hbox{(curvature map)},
\end{eqnarray}
be closable linear operators such that $D_0 \subset D$, $A(D_0)
\subset D$, $S_0(D_0) \subset D(\hat{A})$, $T_0(D_0) \subset D(B)$ and
such that the following intertwining relations hold
\begin{eqnarray}
S_0 A u &=& \hat{A} S_0 u,
\label{Eq:Intertwining1}\\
T_0 A u &=& B T_0 u + L_1 S_0 u + L_0 u,
\label{Eq:Intertwining2}
\end{eqnarray}
for all $u\in D_0$ where $L_1: Y \to Z$, $L_0: X \to Z$ are bounded
linear maps. Furthermore, assume the existence of constants
$a_0,a_1,a_2,\hat{a},b\in \Real$ such that
\begin{equation}
\begin{array}{ll}
(u,Au)_X \leq a_0 \| u \|_X^2 + a_1 \| S_0 u \|_Y^2 + a_2 \| T_0 u \|_Z^2
& \hbox{for all $u\in D_0$,}\\
(c,\hat{A}c)_Y \leq \hat{a} \| c \|_Y^2
& \hbox{for all $c\in D(\hat{A})$,}\\
(w,Bw)_Z \leq b \| w \|_Z^2
& \hbox{for all $w\in D(B)$.}
\end{array}
\label{Eq:Estimates}
\end{equation}

Then,
\begin{enumerate}
\item[(a)] The operator $R_0: D \subset X \to Y \times Z$ defined by
$R_0 u = (S_0 u, T_0 u)$ for all $u\in D$, is closable. The domain
$D(R)$ of its closure $R = \overline{R_0}$ is contained in $D(S) \cap
D(T)$ where $S = \overline{S_0}$, $T = \overline{T_0}$ and $R u =
(Su,Tu)$ for all $u\in D(R)$. The linear space $H = D(R)$, equipped
with the scalar product
\begin{eqnarray}
(u,v)_H &\equiv& (u,v)_X + (R u,R v)_{Y \times Z} \nonumber\\
        &=&      (u,v)_X + (S u,S v)_Y + (T u,T v)_Z\; ,
\nonumber
\end{eqnarray}
for $u,v\in H$ defines a Hilbert space. Furthermore, the restriction
of the operators $S$ and $T$ to $H$ are bounded linear operators from
$H$ to $Y$ and $Z$, respectively.
\item[(b)] If $D_0 \subset H$ is dense in $H$, the operator $A_H:
D(A_H) \subset H \to H$ defined by
\begin{displaymath}
D(A_H) = D_0\; , \qquad A_H u = Au, \qquad u\in D_0\; ,
\end{displaymath}
is densely-defined and quasi-dissipative. In particular, $A_H$ is
closable.
\item[(c)] In addition, assume that $(\lambda - A_H)D_0$ is dense in
$H$ for some sufficiently large $\lambda > 0$. Then, $\overline{A_H}$
generates a strongly continuous semigroup $P: [0,\infty) \to {\cal
L}(H)$ on $H$.
\item[(d)] If in addition $\overline{\hat{A}}$ generates a strongly
continuous semigroup $Q: [0,\infty) \to {\cal L}(Y)$ on $Y$, then
\begin{displaymath}
S P(t) = Q(t) \left. S \right|_H
\end{displaymath}
for all $t \geq 0$. In particular, $\ker \left. S \right|_H$ is
left-invariant under $P(t)$ for all $t\geq 0$.
\end{enumerate}
\end{theorem}

\proof (a) Let $u_n$ be a sequence in $D$ which converges to zero in
$X$ such that $R_0 u_n = (S_0 u_n,T_0 u_n)$ converges to some
$(c,w)\in Y \times Z$. Since $S_0$ and $T_0$ are closable, it follows
that $c=0$ and $w=0$. Therefore, $R_0$ is closable. Next, let $u\in
D(R)$. Then, by definition of the closure, there exists a sequence
$u_n$ in $D$ which converges to $u$ in $X$ such that $R_0 u_n \to R
u$. Since $R_0 u_n = (S_0 u, T_0 u_n)$, it follows that $u\in D(S)
\cap D(T)$ and that $R u = (S u, T u)$. Next, it is clear that $H$ is
linear and that $(.,.)_H$ defines a scalar product on $H$. The
completeness of $H$ is an immediate consequence of the closedness of
$R$. Finally, the fact that $\left. S \right|_H$ and $\left. T
\right|_H$ are bounded is clear.

(b) According to the assumptions, $A(D_0) \subset H$, so $A_H$ is a
well-defined, linear operator. Next, let $u\in D_0$. Using the
intertwining relations (\ref{Eq:Intertwining1},\ref{Eq:Intertwining2})
and the estimates (\ref{Eq:Estimates}) we obtain
\begin{eqnarray}
(u, A_H u)_H &=& (u,A u)_X + (S_0 u, S_0 A u)_Y + (T_0 u, T_0 A u)_Z
\nonumber\\
 &=& (u,A u)_X + (S_0 u, \hat{A}S_0 u)_Y 
  + (T_0 u, B T_0 u + L_1 S_0 u + L_0 u)_Z
\nonumber\\
 &\leq& a_0\| u \|_X^2 + (a_1 + \hat{a}) \| S_0 u \|_Y^2 
 + (a_2 + b) \| T_0 u \|_Z^2
\nonumber\\
 &+& \| L_1 \| \| T_0 u \|_Z \| S_0 u \|_Y 
   + \| L_0 \| \| T_0 u \|_Z \| u \|_X
\nonumber\\
 &\leq& K \| u \|_H^2
\nonumber
\end{eqnarray}
for a sufficiently large constant $K > 0$ which is independent of
$u$. Since $A_H$ is densely-defined and quasi-dissipative it is
closable\footnote{See Theorem 4.5(c) in chapter 1 of
Ref. \cite{Pazy-Book} for a proof.}.

(c) This is a direct consequence of the Lumer-Phillips
theorem\footnote{See Theorem 5.7 in Ref. \cite{hB05} for a proof.}.

(d) For this, we first notice that $\hat{A}$ is closable since
according to our assumptions it is densely-defined and
quasi-dissipative. In a next step we show that
\begin{equation}
S\overline{A}_H u = \overline{\hat{A}} S u
\end{equation}
for all $u\in D(\overline{A_H})$. Let $u\in D(\overline{A_H})$. By
definition of the closure, there exists a sequence $u_n$ in $D(A_H) =
D_0$ such that $u_n \rightarrow u$ in $H$ and $A_H u_n \rightarrow
\overline{A_H} u$ in $H$. Since for all $n\in\Natural$, $S A_H u_n =
\hat{A} S u_n$ and since $\left. S \right|_{H}: H \to Y$ is bounded so
that $S u_n \rightarrow S u$ in $Y$ and $S A_H u_n \rightarrow
S\overline{A_H} u$ in $Y$, it follows that $S u\in
D(\overline{\hat{A}})$ and that $\overline{\hat{A}} Su =
S\overline{A}_H u$.

Next, let $u\in D(\overline{A_H})$ and define the curve $c: [0,\infty) \to
Y$ by $c(t) = S P(t)u$, $t\geq 0$. Since $\left. S \right|_{H}: H \to
Y$ is linear and bounded, $c$ is differentiable on $(0,\infty)$ and
\begin{displaymath}
\frac{d}{dt} c(t) = S\overline{A}_H P(t) u = \overline{\hat{A}} S P(t) u
 = \overline{\hat{A}} c(t)
\end{displaymath}
for all $t > 0$. Since $c(0) = Su$ it follows by uniqueness of the
Cauchy problem associated to $\overline{\hat{A}}$ that $c(t) = Q(t) S
u$. Therefore, $S P(t) u = Q(t) S u$ for all $u\in
D(\overline{A_H})$. Since $D(\overline{A_H})$ is dense in $H$ and $\left. S
\right|_{H}: H \to Y$ is continuous the statement of the theorem
follows.  \qed

In order to apply this theorem to the linearized vacuum equations, we
define the operators $A$, $\hat{A}$, $B$, $S_0$ and $T_0$ and the
function spaces $X$, $Y$ and $Z$ as follows. We start with the
definition of the constraint evolution vector field $\hat{A}$.

\subsection{The constraint propagation system}

Define $Y = H^1(\Sigma,\Real) \times \{ {\bf M} \in H^1(\Sigma,V) :
N^i M_i \hateq 0 \}$, and define the operator $\hat{A}: D(\hat{A})
\subset Y \to Y$ by
\begin{eqnarray}
&& D(\hat{A}) = \{ (H,{\bf M})\in C^\infty(\bar{\Sigma},\Real) 
  \times C^\infty(\bar{\Sigma},V) : N^i D_i H \hateq 0, N^i M_i \hateq 0 \},
\nonumber\\
&& \hat{A}\left( \begin{array}{c} H \\ {\bf M} \end{array} \right)
  = \left( \begin{array}{c} \dvrg {\bf M} \\ \grad H \end{array} \right),
\nonumber
\end{eqnarray}
for all $(H,{\bf M}) \in D(\hat{A})$, where here and in the following
we use the notation $\dvrg {\bf M} \equiv D^i M_i$ and $(\grad H)_i
\equiv D_i H$.

\begin{lemma}[Well posedness of the constraint propagation system]
\label{Lem:WPCPS}
The operator $\hat{A}$ is a densely-defined, linear operator on $Y$
with the following properties:
\begin{enumerate}
\item[(i)] $\hat{A}$ is dissipative, that is, $(c,\hat{A} c)_Y \leq 0$
for all $c\in D(\hat{A})$.
\item[(ii)] For all $\lambda > 0$
\begin{displaymath}
(\lambda - \hat{A})(D(\hat{A})) = C^\infty(\bar{\Sigma},\Real) 
  \times \{ {\bf M}\in C^\infty(\bar{\Sigma},V) : N^i M_i \hateq 0 \}.
\end{displaymath}
\item[(iii)] $\hat{A}$ is closable and its closure generates a
strongly continuous semigroup $Q: [0,\infty) \to {\cal L}(Y)$.
\end{enumerate}
\end{lemma}

\proof
First, it is clear that $\hat{A}$ is linear and that $D(\hat{A})$ is
dense in $Y$.

(i) Let $c = (H,{\bf M})\in D(\hat{A})$. Using integration by parts we find
\begin{eqnarray}
(c,\hat{A}c)_Y &=& \int\limits_\Sigma \left[ H D^i M_i 
 + (D^j H)(D_j D^i M_i) + M^i D_i H + (D^j M^i)(D_j D_i H) \right] d^3 x
\nonumber\\
 &=& \int\limits_\Sigma D^i \left[ H M_i + (D^j H)(D_j M_i) \right] d^3 x
\nonumber\\
 &=& \int\limits_{T^2} \left[ H M_x + (D^j H)(D_j M_x) \right]_{x=0}^1 dy dz.
\nonumber
\end{eqnarray}
Since $M_x \hateq D_x H \hateq 0$ the boundary integral vanishes
and it follows that $\hat{A}$ is dissipative.

(ii) This follows using general theorems \cite{pS96a} about symmetric
linear operators with maximal dissipative boundary
conditions. However, in this particular case, it is not difficult to
give a direct proof: Let $\lambda > 0$, and let $F\in
C^\infty(\bar{\Sigma},\Real)$ and ${\bf G}\in
C^\infty(\bar{\Sigma},V)$ with $N^i G_i \hateq 0$. We want to find
$(H,{\bf M})\in D(\hat{A})$ such that
\begin{eqnarray}
\lambda H - D^i M_i &=& F,
\label{Eq:lambdaH}\\
\lambda M_i - D_i H &=& G_i\; .
\label{Eq:lambdaM}
\end{eqnarray}
Eqs. (\ref{Eq:lambdaH},\ref{Eq:lambdaM}) and $(H,{\bf M})\in D(\hat{A})$
imply the Neumann boundary-value problem
\begin{eqnarray}
&& (\lambda^2 - D^i D_i)H = \lambda F + D^i G_i\; ,
\nonumber\\
&& N^i D_i H \hateq 0,
\nonumber
\end{eqnarray}
which has a unique solution $H\in C^\infty(\bar{\Sigma},\Real)$.
Setting ${\bf M} = \lambda^{-1}( {\bf G} + \grad H ) \in
C^\infty(\bar{\Sigma},V)$ it follows that $(H,{\bf M})\in D(\hat{A})$
satisfies $(\lambda-\hat{A})(H,{\bf M}) = (F,{\bf G})$.

(iii) Since $C^\infty(\bar{\Sigma},\Real) \times \{ {\bf M}\in
C^\infty(\bar{\Sigma},V) : N^i M_i \hateq 0 \}$ is dense in $Y$ this
follows from (i) and (ii) and the Lumer-Phillips theorem\footnote{See
Theorem 5.7 in Ref. \cite{hB05} for a proof.}.
\qed

\subsection{The Weyl propagation system}
\label{SubSect:WeylPS}

Next, we analyze the propagation of the Weyl curvature. As discussed
in Sect. \ref{Sect:MainIdeas} we describe the evolution of the Weyl
tensor by the system constructed in \cite{hFgN99}. For this, define $Z
= L^2(\Sigma,S) \times L^2(\Sigma,S)$, and let $B: D(B) \subset Z \to
Z$ be the densely-defined linear operator
\begin{eqnarray}
&& D(B) = \{ ({\bf E},{\bf B})\in C^\infty(\bar{\Sigma},S) \times C^\infty(\bar{\Sigma},S) : 
  \Psi_0[{\bf E},{\bf B}] \hateq c\bar{\Psi}_4[{\bf E},{\bf B}] \},
\nonumber\\
&& B\left( \begin{array}{c} {\bf E} \\ {\bf B} \end{array} \right)
  = \left( \begin{array}{rr} 
   -(\curl{\bf B})_{ij} + N_{(i}\varepsilon_{j)}{}^{kl} N_k (\dvrg{\bf B})_l\\ 
    (\curl{\bf E})_{ij} - N_{(i}\varepsilon_{j)}{}^{kl} N_k (\dvrg{\bf E})_l 
 \end{array} \right),
\nonumber
\end{eqnarray}
for all $({\bf E},{\bf B}) \in D(B)$. Here, $c$ is a complex constant
of magnitude smaller or equal than one and the maps $({\bf E},{\bf B})
\mapsto \Psi_{0}[{\bf E},{\bf B}]$ and $({\bf E},{\bf B}) \mapsto
\bar{\Psi}_{4}[{\bf E},{\bf B}]$ are defined by
\begin{eqnarray}
\Psi_0[{\bf E},{\bf B}]
 &=& \left[ E_{ij} - \epsilon_{kli} N^k B^l{}_j \right] m^i m^j,
\label{Eq:Psi0}\\
\bar{\Psi}_4[{\bf E},{\bf B}] 
 &=& \left[ E_{ij} + \epsilon_{kli} N^k B^l{}_j \right] m^i m^j.
\label{Eq:Psi4}
\end{eqnarray}
As shown in \cite{hFgN99}, the operator $B$ is symmetric with
maximally dissipative boundary conditions. The following lemma is a
consequence of general theorems \cite{pS96a}.

\begin{lemma}[Well posedness of the Weyl propagation system]
\label{Lem:WPWPS}
The operator $B$ is a densely-defined, linear operator on $Z$ with the
following properties:
\begin{enumerate}
\item[(i)] $B$ is dissipative.
\item[(ii)] For all $\lambda > 0$
\begin{displaymath}
(\lambda - B)(D(B)) = C^\infty(\bar{\Sigma},S) \times C^\infty(\bar{\Sigma},S).
\end{displaymath}
\item[(iii)] $B$ is closable and its closure generates a
strongly continuous semigroup on $Z$.
\end{enumerate}
\end{lemma}

\subsection{The main evolution system}

Next, we turn our attention to the main evolution system. Using the
gauge-fixing projection operator $P^{(GF)} \equiv P_0^{(GF)}$
introduced in the previous section we define the Hilbert space $X =
H^1(\Sigma,S) \times P^{(GF)}L^2(\Sigma,S)$ and the dense subspace
\begin{displaymath}
D = \{ ({\bf h},{\bf k}) \in C^\infty(\bar{\Sigma},S) 
 \times P^{(GF)}C^\infty(\bar{\Sigma},S) : N^i (\M {\bf k})_i \hateq 0 \}.
\end{displaymath}
In terms of the linearized Ricci operator $\Ric: H^2(\Sigma,S) \to
L^2(\Sigma,S)$ defined by
\begin{displaymath}
(\Ric {\bf h})_{ij} = D_{(i} D^k h_{j)k} 
 -\frac{1}{2}\left( D^k D_k h_{ij} + \delta^{kl} D_i D_j h_{kl} \right),
\qquad {\bf h}\in H^2(\Sigma,S),
\end{displaymath}
the main evolution vector field is given by $A: D(A) \subset X \to
X$ where $D(A) = D$ and
\begin{displaymath}
A\left( \begin{array}{c} {\bf h} \\ {\bf k} \end{array} \right)
  = \left( \begin{array}{c} -2{\bf k} \\ 
  P^{(GF)}\Ric {\bf h} \end{array} \right),
\end{displaymath}
for all $({\bf h},{\bf k}) \in D$. 

Next, we define the constraint map $S_0 : D \subset X \to Y$ and the
curvature map $T_0 : D \subset X \to Z$ by
\begin{equation}
S_0\left( \begin{array}{c} {\bf h} \\ {\bf k} \end{array} \right)
  = \left( \begin{array}{c} \Ham {\bf h} \\ \M {\bf k} \end{array} \right),
\qquad
T_0\left( \begin{array}{c} {\bf h} \\ {\bf k} \end{array} \right)
  = \left( \begin{array}{c} [\Ric {\bf h}]^{TF} \\ 
           \curl {\bf k} \end{array} \right),
\qquad \left( \begin{array}{c} {\bf h} \\ {\bf k} \end{array} \right) \in D,
\label{Eq:DefCCMaps}
\end{equation}
where the linearized Hamiltonian operator $\Ham: H^2(\Sigma,S) \to
L^2(\Sigma,\Real)$ is defined by
\begin{displaymath}
\Ham {\bf h} = \frac{1}{2}\delta^{ij}(\Ric {\bf h})_{ij} 
 = \frac{1}{2}\left( D^i D^j h_{ij} - \delta^{ij} D^k D_k h_{ij} \right),
\qquad {\bf h}\in H^2(\Sigma,S).
\end{displaymath}

Finally, we define the subspace $D_0$ of $D$ as the subspace of $X$
consisting of smooth fields satisfying the boundary conditions. More
precisely,
\begin{eqnarray}
D_0 &=& \{ ({\bf h},{\bf k}) \in C^\infty(\bar{\Sigma},S) \times
 P^{(GF)}C^\infty(\bar{\Sigma},S) : \nonumber\\ 
&& \qquad N^i (\M {\bf k})_i \hateq 0, \quad 
 N^i D_i(\Ham {\bf h}) \hateq 0, \quad 
  \Psi_0[ T_0({\bf h},{\bf k}) ] \hateq c\bar{\Psi}_4[ T_0({\bf h},{\bf k}) ]
 \},
 \nonumber
\end{eqnarray}
where $\Psi_0$ and $\bar{\Psi}_4$ are the maps defined in
Eqs. (\ref{Eq:Psi0},\ref{Eq:Psi4}) and $c$ is the complex constant
appearing in the definition of $D(B)$. For the proof below, we require
that $c \neq -1$ which excludes the case of a ``conducting boundary''
described by $m^i m^j E_{ij} \hateq 0$.

In the next Lemma and Propositions \ref{Prop:DenseDomain} and
\ref{Prop:DenseRange} below we show that the assumptions of Theorem
\ref{Thm:LinConstrainedEvolutionCurvMap} are satisfied. This leads to
the main result in Theorem \ref{Thm:Main}.

\begin{lemma}
\label{Lem:BasicProperties}
\begin{enumerate}
\item[(i)] The operators $S_0$ and $T_0$ are closable.
\item[(ii)] $A(D_0) \subset D$, $S_0(D_0) \subset D(\hat{A})$, $T_0(D_0) \subset D(B)$ and
\begin{eqnarray}
S_0 A u &=& \hat{A} S_0 u,
\nonumber\\
T_0 A u &=& B T_0 u + L_1 S_0 u,
\nonumber
\end{eqnarray}
for all $u\in D_0$ where $L_1: Y \to Z$ is the bounded linear operator
defined by
\begin{displaymath}
L_1\left( \begin{array}{c} H \\ {\bf M} \end{array} \right)
 = \left( \begin{array}{c} \frac{1}{2} D_{(i} M_{j)} 
 - \frac{1}{6} \delta_{ij} D^k M_k 
 + \frac{1}{2} N_{(i} \varepsilon_{j)}{}^{kl} N_k (\curl {\bf M})_l \\
 \frac{1}{3} N_{(i} \varepsilon_{j)}{}^{kl} N_k D_l H \end{array} \right).
\end{displaymath}
for all $(H,{\bf M}) \in Y$.
\item[(iii)] There are constants $a_0, a_1, a_2\in \Real$ such that
\begin{displaymath}
(u,Au)_X \leq a_0 \| u \|_X^2 + a_1 \| S_0 u \|_Y^2 + a_2 \| T_0 u \|_Z^2
\end{displaymath}
for all $u\in D_0$.
\end{enumerate}
\end{lemma}

\proof See Appendix \ref{App:Proofs}.
\qed

Following Theorem \ref{Thm:LinConstrainedEvolutionCurvMap} we denote
by $R$ the closure of the operator $R_0: D \subset X \to Y \times Z$
defined by $R_0 u = (S_0 u, T_0 u)$ for all $u\in D$ and define the
Hilbert space $H = D(R)$ equipped with the scalar product
\begin{displaymath}
(u,v)_H \equiv (u,v)_X + (S u,S v)_Y + (T u,T v)_Z
\end{displaymath}
for $u,v\in H$ and define the linear operator $A_H: D_0 \subset H \to
H$ by $A_H u = Au$, for $u\in D_0$. It remains to prove the following
two propositions:

\begin{proposition}
\label{Prop:DenseDomain}
$D_0 \subset H$ is dense.
\end{proposition}

\begin{proposition}
\label{Prop:DenseRange}
$(\lambda - A)(D_0) = D$ for all $\lambda > 0$.
\end{proposition}

For the proof of Proposition \ref{Prop:DenseDomain}, we rely on the
following two lemmas which are proven in Appendix \ref{App:Proofs}.

\begin{lemma}
\label{Lem:h}
Given $\varepsilon > 0$ and $G\in C^\infty(\partial\Sigma,\Real)$
there exists ${\bf h}\in C^\infty(\bar{\Sigma},S)$ such that
\begin{displaymath}
N^i D_i(\Ham {\bf h})\hateq G
\end{displaymath}
and
\begin{displaymath}
\| {\bf h} \|_{H^1(\Sigma,S)} < \varepsilon, \qquad
\| \Ric {\bf h} \|_{L^2(\Sigma,S)} < \varepsilon, \qquad
\| \Ham {\bf h} \|_{H^1(\Sigma,\Real)} < \varepsilon.
\end{displaymath}
\end{lemma}

\begin{lemma}
\label{Lem:k}
Given $\varepsilon > 0$ and $q\in C^\infty(\partial\Sigma,\Complex)$
there exists ${\bf k}\in C^\infty(\bar{\Sigma},S)$ such that
\begin{displaymath}
N^i(\M {\bf k})_i \hateq 0, \qquad
(\curl {\bf k})_{ij} m^i m^j \hateq q
\end{displaymath}
and
\begin{displaymath}
\| {\bf k} \|_{L^2(\Sigma,S)} < \varepsilon, \qquad
\| \curl {\bf k} \|_{L^2(\Sigma,S)} < \varepsilon, \qquad
\| \M {\bf k} \|_{H^1(\Sigma,V)} < \varepsilon.
\end{displaymath}
\end{lemma}

\proofof{Proposition \ref{Prop:DenseDomain}} 
Let $u = ({\bf h},{\bf k})\in H$ and $\varepsilon > 0$ be given. We
have to show that there exists $\hat{u}\in D_0$ such that $\| \hat{u}
- u \|_H < \varepsilon$. In order to do so, we first notice that $D$
is dense in $H$ and pick $\bar{u} = (\bar{\bf h},\bar{\bf k})\in D$
such that
\begin{displaymath}
\| \bar{u} - u \|_H^2 < \frac{\varepsilon^2}{2}\; . 
\end{displaymath}
Next, we use Lemma \ref{Lem:h} and choose $\tilde{\bf h}\in
C^\infty(\bar{\Sigma},S)$ such that $N^i D_i(\Ham \tilde{\bf h})
\hateq -N^i D_i(\Ham \bar{\bf h})$ and
\begin{displaymath}
\| \tilde{\bf h} \|_{H^1(\Sigma,S)}^2 < \frac{\varepsilon^2}{12}\; ,
\qquad
\| \Ric \tilde{\bf h} \|_{L^2(\Sigma,S)}^2 < \frac{\varepsilon^2}{12}\; ,
\qquad
\| \Ham \tilde{\bf h} \|_{H^1(\Sigma,\Real)}^2 < \frac{\varepsilon^2}{12}\; .
\end{displaymath}
Next, we notice that the boundary condition $\Psi_0[{\bf E},{\bf B}]
\hateq c\bar{\Psi}_4[{\bf E},{\bf B}]$ is equivalent to
\begin{displaymath}
B_{ij} m^i m^j \hateq \frac{1}{i}\frac{1-c}{1+c} E_{ij} m^i m^j.
\end{displaymath}
Therefore, we use Lemma \ref{Lem:k} and pick $\tilde{\bf k}\in
C^\infty(\bar{\Sigma},S)$ such that $\M\tilde{\bf k} \hateq 0$,
\begin{displaymath}
\curl[\bar{\bf k} + \tilde{\bf k}]_{ij} m^i m^j
 \hateq 
 \frac{1}{i}\frac{1-c}{1+c}[\Ric(\bar{\bf h}+\tilde{\bf h})]_{ij} m^i m^j
\end{displaymath}
and
\begin{displaymath}
\| \tilde{\bf k} \|_{L^2(\Sigma,S)}^2 < \frac{\varepsilon^2}{12}\; ,
\qquad
\| \curl\tilde{\bf k} \|_{L^2(\Sigma,S)}^2 < \frac{\varepsilon^2}{12}\; ,
\qquad
\| \M\tilde{\bf k} \|_{H^1(\Sigma,V)}^2 < \frac{\varepsilon^2}{12}\; .
\end{displaymath}
Finally, set $\hat{u} = (\bar{\bf h} + \tilde{\bf h}, \bar{\bf k} +
P^{(GF)}\tilde{\bf k})$. Noticing that $\M P^{(GF)}\tilde{\bf k} =
\M\tilde{\bf k}$ and $\curl P^{(GF)}\tilde{\bf k} = \curl\tilde{\bf
k}$ it follows that $\hat{u}\in D_0$ and that
\begin{eqnarray}
\| \hat{u}-u \|_H^2 &\leq& \| \hat{u}-\bar{u} \|_H^2 + \| \bar{u}-u \|_H^2
\nonumber\\
 &=& \| \tilde{\bf h} \|_{H^1(\Sigma,S)}^2 
 + \| P^{(GF)}\tilde{\bf k} \|_{L^2(\Sigma,S)}^2
 + \| \Ham \tilde{\bf h} \|_{H^1(\Sigma,\Real)}^2 
 + \| \M\tilde{\bf k} \|_{H^1(\Sigma,V)}^2
\nonumber\\
&+& \| [\Ric \tilde{\bf h}]^{TF} \|_{L^2(\Sigma,S)}^2
 + \| \curl\tilde{\bf k} \|_{L^2(\Sigma,S)}^2
 + \| \bar{u}-u \|_H^2
\nonumber\\
 & < & \varepsilon^2,
\nonumber
\end{eqnarray}
where we have used the fact that $P^{(GF)}$ is a projector in the last
step.
\qed

\proofof{Proposition \ref{Prop:DenseRange}} 
Let $\lambda > 0$ and $F\in D$. In the following, we construct $u =
({\bf h},{\bf k})\in D_0$ such that $(\lambda - A)u = F$. For this, we
first use Lemma \ref{Lem:WPCPS}(ii) and find $(H,{\bf M})\in
D(\hat{A})$ such that
\begin{equation}
(\lambda - \hat{A})\left( \begin{array}{c} H \\ {\bf M} \end{array} \right)
 = S_0 F.
\label{Eq:ConstraintSolve}
\end{equation}
Next, using Lemma \ref{Lem:WPWPS}(ii) we find $({\bf E},{\bf B})\in
D(B)$ such that
\begin{equation}
(\lambda - B)\left( \begin{array}{c} {\bf E} \\ {\bf B} \end{array} \right)
 = T_0 F + L_1\left( \begin{array}{c} H \\ {\bf M} \end{array} \right).
\label{Eq:WeylSolve}
\end{equation}
In a next step, we define the auxiliary variables
\begin{displaymath}
{\bf Q} = \dvrg {\bf E} - \frac{1}{3}\grad H, \qquad
{\bf P} = \dvrg {\bf B} + \frac{1}{2}\curl {\bf M},
\end{displaymath}
and prove that they vanish. Multiplying ${\bf Q}$ and ${\bf P}$ by
$\lambda$ and using Eqs. (\ref{Eq:ConstraintSolve},\ref{Eq:WeylSolve})
we obtain the system
\begin{displaymath}
\lambda\left( \begin{array}{c} Q_x \\ Q_A \\ P_x \\ P_A \end{array} \right)
 = \left( \begin{array}{r}
    -\varepsilon^{AB} D_A P_B \\ -\frac{1}{2}\varepsilon_A{}^B D_B P_x \\
     \varepsilon^{AB} D_A Q_B \\  \frac{1}{2}\varepsilon_A{}^B D_B Q_x \\
   \end{array} \right),
\end{displaymath}
where $A,B$ run over $y$ and $z$. Multiplying both sides from the left
with $(Q^x,2Q^A,P^x,2P^A)$ and integrating over $\Sigma$, it follows
that ${\bf P} = {\bf Q} = 0$, as claimed.

Next, we set
\begin{displaymath}
u = \left( \begin{array}{c} {\bf h} \\ {\bf k} \end{array} \right)
  = \frac{1}{\lambda}\left[ F 
 + \left(\begin{array}{c} -2{\bf k} \\ 
   P^{(GF)}\left( {\bf E} + \frac{2}{3}{\bf \delta} H \right) \end{array} 
 \right) \right].
\end{displaymath}
Since $N^i D_i H \hateq 0$ and ${\bf Q} = 0$ it follows that
$N^i\M({\bf E} + \frac{2}{3}{\bf \delta} H)_i \hateq -N^i\dvrg {\bf
E}_i + \frac{4}{3} N^i D_i H \hateq 0$. Therefore, it follows that
$u\in D$. Next, using Eq. (\ref{Eq:ConstraintSolve}) and ${\bf Q} = 0$
we find
\begin{displaymath}
\lambda\left[\left( \begin{array}{c} H \\ {\bf M} \end{array} \right) 
 - S_0 u\right] 
 = \left( \begin{array}{c} \dvrg {\bf M} - \dvrg\M {\bf k} \\ 0 \end{array} 
 \right)
\end{displaymath}
which proves that $(H,{\bf M}) = S_0 u$. Finally, Eq. (\ref{Eq:WeylSolve}),
${\bf P} = {\bf Q} = 0$ and ${\bf M} = \M {\bf k}$ yield
\begin{displaymath}
\lambda\left[\left( \begin{array}{c} {\bf E} \\ {\bf B} \end{array} \right) 
 - T_0 u\right] 
 = \left( \begin{array}{c} -\curl {\bf B} + \curl^2 {\bf k} \\ 0 \end{array} 
\right)
\end{displaymath}
which shows that $({\bf E},{\bf B}) = T_0 u$. Therefore, $u\in D_0$ and
$(\lambda-A)u = F$.
\qed

To summarize, we have shown:

\begin{theorem}[Main result]
\label{Thm:Main}
Let $c\in \Complex$ be such that $|c| \leq 1$ and $c\neq -1$. The
operator $A: D_0 \subset H \to H$
(cf. Eq. (\ref{Eq:EvolutionVectorField})) describing the linearized
Einstein evolution equations is closable and its closure generates a
strongly continuous semigroup $P: [0,\infty) \to {\cal L}(H)$ on
$H$. In particular, given initial data $u_0 = ({\bf h},{\bf k})\in
C^\infty(\bar{\Sigma},S) \times C^\infty(\bar{\Sigma},S)$ satisfying
the gauge condition $P^{(GF)} {\bf k} = {\bf k}$ and the compatibility
conditions
\begin{eqnarray}
&& N^i(D_i k - D^j k_{ij}) \hateq 0,
\nonumber\\
&& N^i D_i\left( D^k D^l h_{kl} - \delta^{kl} D^j D_j h_{kl} \right) \hateq 0,
\nonumber\\
&& \Psi_0[T_0({\bf h},{\bf k})] \hateq c \bar{\Psi}_4[T_0({\bf h},{\bf k})],
\nonumber
\end{eqnarray}
where $\Psi_0$ and $\bar{\Psi}_4$ are the maps defined in
Eqs. (\ref{Eq:Psi0},\ref{Eq:Psi4}) and $T_0$ is the curvature map
defined in Eq. (\ref{Eq:DefCCMaps}), the curve $u: [0,\infty) \to H, t
\mapsto u(t) \equiv P(t)u_0$ is continuous and differentiable on
$(0,\infty)$ and is the unique solution of the abstract Cauchy problem
(\ref{Eq:CauchyProblem}). Furthermore, if $u_0$ satisfies the
constraints, i,e, if $S_0 u_0 = 0$, the constraints are satisfied for
all $t > 0$, i.e. $S u(t) = 0$ for all $t > 0$.
\end{theorem}

\noindent
{\bf Remarks}:
\begin{enumerate}
\item For each $t > 0$ the solution $u(t)$ lies not only in $H$ but in
the domain $D(\overline{A_H})$ of the closure $\overline{A_H}$ of $A_H$. We do
not intend to give an explicit representation of this space in this
article.
\item It should be possible to show that the solution is smooth
provided extra compatibility conditions are required for the initial
data.
\item Using Duhamel's principle\footnote{See, for example, Corollary
2.11 in chapter 4 of Ref. \cite{Pazy-Book}.}, it is not difficult to
generalize the theorem to inhomogeneous boundary conditions and to
the presence of source terms in the evolution equation.
\end{enumerate}

\section{Conclusions}
\label{Sect:Conclusions}

In this article, we have discussed some new ideas for tackling the
problem of obtaining a well posed IBVP for metric formulations of
Einstein's field equations. These ideas go beyond casting the
evolution equations into symmetric hyperbolic form with maximally
dissipative boundary conditions which, except in some rather
restricted situations \cite{bSbSjW02,bSjW03,gCjPoRoSmT03,cGjM04b}, do
not seem to be flexible enough to be made compatible with the
constraints for metric-based formulations. In particular, we have
analyzed a gauge condition for the lapse which is obtained by
minimizing a functional representing the norm squared of the time
derivative of the extrinsic curvature. This leads to a fourth order
elliptic boundary value problem. We have shown that coupling this
elliptic problem to the Einstein evolution equations in the $3+1$
decomposition introduced in \cite{jY79} with a zero shift and imposing
constraint-preserving boundary conditions controlling the Weyl scalar
$\Psi_0$ leads to a well posed IBVP in the linearized regime. Despite
the fact that the results so far have only been obtained in the weak
field regime, there are several interesting points on which we would
like to comment.

First, it is known that when the lapse is frozen or densitized, the
evolution equations we have analyzed are only weakly hyperbolic
\cite{KreissOrtiz-2002,gNoOoR04}. For such systems there are examples
of solutions with frequency-dependent exponentially growing modes
\cite{KL-Book,KreissOrtiz-2002,gNoOoR04}. It is therefore a priori not
clear that a well posed IBVP can be obtained for the Einstein
evolution equations in \cite{jY79}. On the other hand, as discussed in
Sect. \ref{Sect:MainIdeas}, even though these equations are weakly
hyperbolic, they induce a ``nice'' evolution for the constraint and
curvature fields. The well posed initial-boundary value formulation
presented in this article is based on this particular property of the
evolution equations and on the implementation of a fourth order
elliptic gauge condition for the lapse instead of a frozen or
densitized lapse. A well posed elliptic-hyperbolic formulation for
the full nonlinear Einstein equations without boundaries was given in
\cite{lAvM03}.

The next point is related to the discretization of the problem. First,
our formulation requires solving a fourth order elliptic equation at
each (or each few) timesteps which might be computationally
expensive. In addition to that, since our well posedness proof relies
on the propagation of the constraints, it is a priori not clear that a
``naive'' discretization of the problem will lead to a stable and
convergent scheme. It might be the case that one has to choose very
special discretizations techniques such that the estimates in the
continuum case can be mimicked at the discrete level. This may require
some constraint projection mechanism.

Next, we would like to remark that our approach is quite general and
should work for any metric formulation of the Einstein equations for
which the constraint propagation system can be cast into symmetric
hyperbolic form and for which sufficient regularity for the constraint
fields can be shown. In particular, imposing the same gauge condition
as in this article, we expect that a well posed IBVP can also be
derived for families of linearized symmetric hyperbolic first order
formulations \cite{sFoR96,aAjY99} or linearized mixed first-order
second-order hyperbolic formulations \cite{gNoOoR04,cGjM04b} which
might be more suitable for numerical discretization.

Finally, the question remains as to what other gauge conditions may
work in our approach. In particular, the geometrical meaning of the
boundary surface obtained (as embedded in the spacetime constructed)
must be clarified. This issue is likely related to the choice of the
shift vector. In \cite{sD04} a functional $J(\alpha,\beta^i)$ of the
lapse $\alpha$ and the shift vector $\beta^i$ was introduced in order
to construct approximate Killing fields for a given data set. For zero
shift our functional $I_{GR}[\alpha]$ is closely related to $J$. It
should be interesting to investigate gauge conditions that are
obtained by varying $J$ with respect to lapse and shift.

These questions, as well as the generalization to linearizations about
more general spacetimes including inner excision boundaries, will be
considered in future work.

\section{Acknowledgments}

It is a pleasure to thank H. Beyer, G. Calabrese, M. Holst, L. Lehner,
H. Pfeiffer, O. Reula and M. Tiglio for helpful discussions and
suggestions and D. Reynolds for reading the manuscript. This research
was supported in part by NSF grant PHY-0099568, by a grant from the
Sherman Fairchild Foundation to Caltech and by NSF DMS Award 0411723
to UCSD.

\appendix
\section{Compact perturbations of coercive operators}
\label{App:Fred}

In this appendix we state the following theorem which is a summary of
well-known results.

\begin{theorem}
\label{Thm:Fred}
Let $X$ be a reflexive Banach space, and let $L: X \to X^*$ be a
linear, bounded coercive operator, i.e. $L\in {\cal L}(X,X^*)$ and
there exists $\delta > 0$ such that $L(u)(u) \geq \delta\| u \|^2$ for
all $u\in X$. Furthermore, let $K\in {\cal L}(X,X^*)$ be a compact
linear operator and set $M:=L + K$. Finally, let $M^t: X \to X^*$
denote the bounded linear operator defined by $(M^t u)(v) = (M v)(u)$
for all $u,v\in X$. Then,
\begin{enumerate}
\item[(i)] $L: X \to X^*$ is invertible with bounded inverse, and
$\| L^{-1} \| \leq \delta^{-1}$.
\item[(ii)] $\ker M$ and $\ker M^t$ are finite dimensional and have
equal dimensions.
\item[(iii)] $\ran M = (\ker M^t)^\perp$, where $(\ker M^t)^\perp$
denotes the annihilator of $\ker M^t$.
\end{enumerate}
\end{theorem}

\proof 
Let $I: X \to X^{**}$ denote the map defined by $(Iu)(\omega) =
\omega(u)$ for all $u\in X$, $\omega\in X^*$. Since $X$ is reflexive,
$I$ is an isometric isomorphism. Next, let $M^*\in {\cal
L}(X^{**},X^*)$ denote the (Banach space) adjoint of $M$, defined by
$(M^* u^{**})(v) = u^{**}(Mv)$ for all $v\in X$, $u^{**}\in
X^{**}$. Then, $M^t = M^* \circ I$. For the following, we use the
formula\footnote{See, for instance, \cite{Kato-Book} Sect. III.}
\begin{equation}
\overline{\ran A} = (\ker A^t)^\perp
\label{Eq:Ranker}
\end{equation}
which holds for any bounded linear operator $A: X \to X^*$ with $A^t =
A^* \circ I$ on a reflexive Banach space $X$.

(i) The coercivity of $L$ implies that
\begin{equation}
\| L u \| \geq \delta \| u \|
\label{Eq:LIneq}
\end{equation}
and
\begin{equation}
\| L^t u \| \geq \delta \| u \|
\label{Eq:LtIneq}
\end{equation}
for all $u\in X$. The first inequality implies that $L$ has trivial
kernel and closed range, the second inequality implies that $L^t$ has
trivial kernel. Eq. (\ref{Eq:Ranker}) then implies that $L$ is
bijective.  $\| L^{-1} \| \leq \delta^{-1}$ now follows from the
inequality (\ref{Eq:LIneq}).

(ii) It follows from (i) that $M = L + K$ is Fredholm with the same
index as $L$ which is zero\footnote{See, for instance,
\cite{Taylor-Book}, Corollaries A.7.2 and A.7.5.}. In particular, $M$
has closed range, and $\dim\ker M = \dim\ker M^t < \infty$.

(iii) Since $M$ has closed range,
\begin{equation}
\ran M = \overline{\ran M} = (\ker M^t)^\perp.
\end{equation}

\qed

\section{Proof of Lemma \ref{Lem:BasicProperties},\ref{Lem:h} and \ref{Lem:k}}
\label{App:Proofs}

\proofof{Lemma \ref{Lem:BasicProperties}}
(i) Suppose $u_n$ is a sequence in $D$ which converges to zero in $X$
and such that $Su_n$ converges to $v\in Y$. We have to show that
$v=0$. In order to see this, write $u_n = ({\bf h}^{(n)},{\bf
k}^{(n)})$ and $v = (H,{\bf M})$. Then, ${\bf h}^{(n)} \rightarrow 0$
in $H^1(\Sigma,S)$, $\Ham {\bf h}^{(n)} \rightarrow H$ in
$H^1(\Sigma,\Real)$ and ${\bf k}^{(n)} \rightarrow 0$ in
$L^2(\Sigma,S)$, $\M {\bf k}^{(n)}] \rightarrow {\bf M}$ in
$H^1(\Sigma,V)$.

Next, take a test function $\varphi\in C_0^\infty(\Sigma,\Real)$. Then,
\begin{eqnarray}
(\varphi,H)_{L^2(\Sigma,\Real)}
 &=& \lim\limits_{n\rightarrow \infty}
 (\varphi, \Ham {\bf h}^{(n)})_{L^2(\Sigma,\Real)}
\nonumber\\
 &=& \lim\limits_{n\rightarrow \infty}
\frac{1}{2} (D^i\varphi, \delta^{kl} D_i h_{kl}^{(n)} 
 - D^j h_{ij}^{(n)})_{L^2(\Sigma,V)}
\nonumber\\
 &=& 0.
\nonumber
\end{eqnarray}
Since this holds for all $\varphi\in C_0^\infty(\Sigma,\Real)$ it
follows that $H = 0$. Similarly, let ${\bf \psi}\in
C_0^\infty(\Sigma,V)$,
\begin{eqnarray}
({\bf \psi},{\bf M})_{L^2(\Sigma,V)}
 &=& \lim\limits_{n\rightarrow \infty} (\psi,\M {\bf k}^{(n)} )_{L^2(\Sigma,V)}
\nonumber\\
 &=& \lim\limits_{n\rightarrow \infty}
 (D^{(i}\psi^{j)}, k_{ij}^{(n)}
 - \delta_{ij}\delta^{kl} k_{kl}^{(n)})_{L^2(\Sigma,S)}
\nonumber\\
 &=& 0,
\nonumber
\end{eqnarray}
which proves that ${\bf M} = 0$. Therefore, $S_0$ is closable. The
proof that $T_0$ is closable is similar.

(ii) This follows by direct verification.

(iii) Let $u = ({\bf h},{\bf k}) \in D_0$. Then,
\begin{eqnarray}
(u,Au)_X &=& ({\bf h},-2{\bf k})_{H^1(\Sigma,S)} 
 + ({\bf k}, P^{(GF)}\Ric {\bf h})_{L^2(\Sigma,S)}
\nonumber\\
 &\leq& \| {\bf h} \|_{H^1(\Sigma,S)}^2 + \| {\bf k} \|_{H^1(\Sigma,S)}^2
 + \frac{1}{2} \| {\bf k} \|_{L^2(\Sigma,S)}^2 
 + \frac{1}{2} \| \Ric {\bf h} \|_{L^2(\Sigma,S)}^2
\nonumber\\
 &\leq& \| {\bf h} \|_{H^1(\Sigma,S)}^2 
 + C_1\left[ \| {\bf k} \|_{L^2(\Sigma,S)}^2
 + \| \curl {\bf k} \|_{L^2(\Sigma,S)}^2 + \| \M {\bf k} \|_{L^2(\Sigma,V)}^2 \right]
\nonumber\\
 &+& C_2\left[ \| [\Ric {\bf h}]^{TF} \|_{L^2(\Sigma,S)}^2 
             + \| \Ham {\bf h} \|_{L^2(\Sigma,\Real)}^2 \right]
\nonumber\\
 &\leq& a_0 \| u \|_X^2 + a_1 \| S_0 u \|_Y^2 + a_2 \| T_0 u \|_Z^2,
\nonumber
\end{eqnarray}
where $C_1$, $C_2$, $a_0$, $a_1$, $a_2$ are constants independent of
$u$ and where we have used Lemma \ref{Lem:KeyEstimate} in the third
step.
\qed

\proofof{Lemma \ref{Lem:h}} 
First, we can assume that $G$ vanishes on one of the boundary
components, say on $x=1$. Otherwise, one can construct ${\bf h}$ as a
superposition of two fields each one satisfying the statement of the
lemma for $G$ vanishing on $x=0$ and $x=1$, respectively. In order to
proceed, let $\psi\in C^\infty( [0,\infty),\Real)$ be such that
\begin{enumerate}
\item[(i)] $\psi'''(0) = 1$,
\item[(ii)] $\psi(x) = 0$ for all $x \geq 2$
\end{enumerate}
and set $\psi_n(x):=n^{-3}\psi(nx)$ for all $n > 2$ and $x \in [0,1]$.
By construction, $\psi_n'''(0) = 1$ and $\psi_n(x) = 0$ for all
$x\in [2/n,1]$. Next, define for each $n > 2$
\begin{displaymath}
h_{ij}^{(n)}(x,y,z) = (2 N_i N_j - \delta_{ij}) \psi_n(x) G(y,z).
\end{displaymath}
Since $\psi_n'''(x) = \psi'''(nx) = 0$ for all $x\in [2/n,1]$ it
follows that
\begin{displaymath}
\| {\bf h}^{(n)} \|_{H^3(\Sigma,S)}^2 \leq \frac{K}{n}
\end{displaymath}
for a sufficiently large constant $K$ independent of $n$. Therefore,
the $H^3$-norm of ${\bf h}^{(n)}$ can be made arbitrarily
small. Furthermore, one finds
\begin{displaymath}
\Ham {\bf h}^{(n)} = \frac{1}{n}\psi''(nx) G(y,z),
\end{displaymath}
so $\left. N^i D_i(\Ham {\bf h}^{(n)})\right|_{x=0} = G$.
\qed

\proofof{Lemma \ref{Lem:k}} 
Similarly to the proof of Lemma \ref{Lem:h} we may assume that $q$
vanishes on $x=1$. Next, let $\Psi\in C^\infty( [0,\infty),\Real)$ be
such that
\begin{enumerate}
\item[(i)] $\Psi(0) = 0$,
\item[(ii)] $\Psi'(0) = 1$
\item[(iii)] $\Psi(x) = 0$ for all $x \geq 2$
\end{enumerate}
and set $\Psi_n(x):=n^{-1}\Psi(nx)$ for all $n > 2$ and $x \in [0,1]$.
By construction, $\Psi_n(0)=0$, $\Psi_n'(0) = 1$ and $\Psi_n(x) = 0$
for all $x\in [2/n,1]$. Next, define for each $n > 2$
\begin{displaymath}
k_{kl}^{(n)}(x,y,z) = -i\Psi_n(x) q(y,z)\bar{m}_k\bar{m}_l + c.c.
\end{displaymath}
where $c.c.$ denotes the complex conjugate of the previous expression.
Since $\Psi_n'(x) = \Psi'(nx) = 0$ for all $x\in [2/n,1]$ it follows
immediately that there is a constant $K > 0$ such that
\begin{displaymath}
\| {\bf k}^{(n)} \|_{H^1(\Sigma,S)}^2 \leq \frac{K}{n}
\end{displaymath}
for all $n > 2$. Therefore, the $H^1$-norm of ${\bf k}^{(n)}$ can be made
arbitrarily small. Next, since $\Psi_n$ depends on $x$ only one finds
\begin{displaymath}
(\M {\bf k}^{(n)})_l = i\Psi_n\bar{m}_l\bar{m}^k D_k q + c.c.,
\end{displaymath}
hence $\left. \M {\bf k}^{(n)} \right|_{x=0} = 0$ and $\| \M {\bf k}^{(n)}
\|_{L^2(\Sigma,V)}$ can be made arbitrarily small. Finally,
\begin{displaymath}
(\curl {\bf k}^{(n)})_{kl} = \Psi'(nx)\bar{m}_k\bar{m}_l q 
 + \Psi_n(x) N_{(k}\bar{m}_{l)}\bar{m}^j D_j q + c.c.
\end{displaymath}
which implies that $\left. (\curl {\bf k}^{(n)})_{kl} m^k m^l
\right|_{x=0} \hateq q$.
\qed

\bibliography{adm}

\end{document}